\documentclass[Proceedings,letterpaper,InsideFigs]{ascelike-new}
\usepackage{amssymb, amsfonts}
\usepackage{booktabs}
\usepackage{xurl}
 
\usepackage{tikz}          
\usepackage{pgfplots}      
\pgfplotsset{compat=1.18}  

\usetikzlibrary{arrows.meta, positioning, calc}
\usepackage{amsthm}
\usepackage{pgfplots}
\usepackage{algorithm}
\usepackage{algorithmic}
\usepackage{xcolor}

\usepackage[utf8]{inputenc}
\usepackage[T1]{fontenc}
\usepackage{lmodern}
\usepackage{graphicx}
\usepackage[style=base,figurename=Fig.,labelfont=bf,labelsep=period]{caption}
\usepackage{subcaption}
\usepackage{amsmath}
\usepackage{newtxtext,newtxmath}
\usepackage[colorlinks=true,citecolor=red,linkcolor=black]{hyperref}
\newtheorem{theorem}{Theorem}
\newtheorem{assumption}{Assumption}
\NameTag{\today}
\begin{document}

\title{Adaptive Kernel Regression for Constrained Route Alignment: Theory and Iterative Data Sharpening}
\author[1]{Shiyin Du}
\author[1]{Yiting Chen}
\author[2,*]{Wenzhi Yang}
\author[3,*]{Qiong Li}
\author[1,*]{Xiaoping Shi}

\affil[1]{University of British Columbia, Department of Computer Science, Mathematics, Physics and Statistics, Kelowna, V1V 1V7, Canada}
\affil[2]{School of Big Data and Statistics, Anhui University, Hefei, 230601, China}
\affil[3]{Guangdong Provincial Key Laboratory of Interdisciplinary Research and Application; Data Science,  Beijing Normal-Hong Kong Baptist University, Zhuhai, 519087, China}
\affil[*]{wzyang@ahu.edu.cn, qiongli@bnbu.edu.cn, xiaoping.shi@ubc.ca}

\maketitle

\begin{abstract}
Route alignment design in surveying and transportation engineering frequently involves fixed waypoint constraints, where a path must precisely traverse specific coordinates. While existing literature primarily relies on geometric optimization or control-theoretic spline frameworks, there is a lack of systematic statistical modeling approaches that balance global smoothness with exact point adherence. This paper proposes an Adaptive Nadaraya-Watson (ANW) kernel regression estimator designed to address the fixed waypoint problem. By incorporating waypoint-specific weight tuning parameters, the ANW estimator decouples global smoothing from local constraint satisfaction, avoiding the ``jagged'' artifacts common in naive local bandwidth-shrinking strategies. To further enhance estimation accuracy, we develop an iterative data sharpening algorithm that systematically reduces bias while maintaining the stability of the kernel framework. We establish the theoretical foundation for the ANW estimator by deriving its asymptotic bias and variance and proving its convergence properties under the internal constraint model. Numerical case studies in 1D and 2D trajectory planning demonstrate that the method effectively balances root mean square error (RMSE) and curvature smoothness. Finally, we validate the practical utility of the framework through empirical applications to railway and highway route planning. In sum, this work provides a stable, theoretically grounded, and computationally efficient solution for complex, constrained alignment design problems.
\end{abstract}


\section{Introduction}
In the route alignment planning of surveying and mapping, constraints often exist. In practical route alignment design problems, precise values are often known at specific points, leading to constraints requiring paths to pass through these given points exactly, we call them fixed waypoints. One of the most common examples of this problem is the control of the starting and ending points of a physical process \cite{kano2011optimal}. Fixed waypoints problem arises in a wide range of transportation and surveying fields. In road and highway design, \citeN{kang2007highway} investigates the use of the “feasible gate” method to replace rigid point-passing constraints when routing through designated key points (or areas) during road construction, enabling routes that satisfy geometric requirements while being more practical. \citeN{maji2017optimization} employs the Path Planner Method to automatically generate highways that traverse designated points and bypass restricted zones, minimize construction costs, and produce routes that are both short and smooth. In railway alignment, \citeN{wan2024railway} classifies areas suitable for railway construction as “feasible islands,” stipulating that railways must traverse these designated zones. Connections are then established between these islands, ultimately determining the shortest possible route. There are also some fixed waypoints problem in other domains, such as maritime and oceanographic applications \cite{zhen2023improved,kytariolou2022ship} and aerospace trajectory planning \cite{mellinger2011minimum,richter2016polynomial}.

These route planning studies involving waypoint constraints primarily rely on computational optimization and algorithmic frameworks to address the requirement of “passing through specific locations,” with minimal consideration given to statistical modeling methods. In addition, existing works design alignment of a road from the perspective of geometry and optimation \cite{doi:10.1061/JSUED2.SUENG-1391}. It is essential to develop a systematic statistical alignment design methods to address fixed waypoints problem. 

In statistical modeling, constrained nonparametric regression aims to estimate a smooth function while enforcing structural restrictions. Among the early studies, \citeN{hall2001nonparametric} studied kernel regression under monotonicity constraints, while \citeN{racine2009constrained} extended the method by using quadratic programming to bind interval constraints into equality constraints. \citeN{henderson2009imposing} further focused on the computability of constraints in practical problems, translating structural constraints in economics into quadratic programming form, thereby achieving the transition from theory to algorithm. Subsequently, the focus of research gradually shifted from general shape constraints to fix waypoints constraints. \citeN{xiong2021reconstruction} proposed the reconstruction method, which generalizes fixed waypoints constraint problems into nonparametric regression problems. Building upon the “reconstruction” concept, \citeN{dontchev2022constrained} extended the fixed waypoints constraint problem to a constrained smoothing problem under optimal control. Parallel to the kernel-based developments, spline-based approaches offered another practical route for incorporating equality and inequality constraints in nonparametric regression \cite{saeidian2021new}. \citeN{sun2000control} formulated smoothing splines within a control-theoretic framework, where the spline is obtained by solving an optimal control problem that minimizes the $L^2$-norm of the control input while keeping the system output close to prescribed interpolation points. \citeN{kano2003b} established a close link between B-splines and linear control systems, formulating interpolation and smoothing splines as optimal control problems that minimize control energy while driving system trajectories through prescribed data points. \citeN{kano2011optimal} developed a quadratic programming framework for equality and inequality constrained splines and demonstrated its use in trajectory planning with fixed waypoints constraints on boundary conditions. These methods are mainly developed using splines or control-based models, rather than kernel regression. And most fixed waypoints constraints represent interpolation problems rather than the fixed waypoint problems that truly correspond to real-world surveying challenges.

Existing research still lacks systematic and specific solutions for nonparametric regression methods that satisfy fixed waypoints constraints. First, in route alignment design scenarios, most work focuses on pure optimization without treating fixed points as non-parametric regression problems. Second, among existing kernel-based methods, few offer simple, interpretable, and easily implementable concrete solutions for fixed waypoint systems. Furthermore, theoretical discussions on the convergence of constrained points remain limited. To address this, we propose an adaptive Nadaraya--Watson kernel regression estimator (ANW) by explicitly incorporating waypoint constraints through waypoint-specific fixed weight tuning parameters (denoted by $\lambda_j$). This approach not only preserves the consistent convergence properties of traditional Nadaraya–Watson kernel regression but also allows the fitted curve to be pulled toward fixed waypoints in a controlled and interpretable way, making it better suited for alignment design tasks, including road and railway route planning, traffic trajectory modeling, and other spatial path design problems.

The remainder of this paper is organized as follows. In the Modeling section, we introduce the internal constraint model and develop the naive bandwidth-shrinking estimator alongside the proposed ANW estimator, providing a comparative analysis of their localized behavior. We then detail the iterative data sharpening procedure for the ANW estimator and provide a formal algorithm for bias reduction. In the Asymptotic Properties section, we establish the theoretical framework for the ANW estimator and derive its asymptotic properties, with all formal mathematical proofs deferred to the Appendix. The Numerical Properties section evaluates the performance and stability of the method through numerical case studies in 1D and 2D trajectory planning. The practical utility of the framework is demonstrated in the Empirical Applications section through its implementation in railway and highway route planning. Finally, the paper concludes with a summary of findings and a discussion of future research directions.

\section{Modelling}

In this study, $X$  denotes the arc length s   $s$ along the route. It represents the position along the route in practical alignment design. The response $Y$ corresponds to a geometric quantity varying along the route, for example elevation, lateral offset, or coordinate components $x(s), y(s)$. Let $(X_1, Y_1), \dots,$ $ (X_n, Y_n)$ be independent and identically distributed $\mathbb{R}^{d+1}$ valued random vectors with $Y$ real valued. We often assume a model of the form $Y_i = m(X_i) + \varepsilon_i$, where $m(x)$ is an unknown, smooth function and $\varepsilon_i$ represents random error. Consider the problem of estimating the regression function,

\[m(x) = \mathbb{E}[Y|X = x],\]
using $(X_1, Y_1), \ldots, (X_n, Y_n)$ without making strong assumptions about its functional form.


Kernel estimators \cite{nadaraya1964estimating,watson1964smooth} provide  an intuitive estimate of $m(x)$, are a
local weighted average of the $Y_i$, given by

\[ \hat{B}_{\mathrm{NW}}(x) =\hat{m}(x) = \frac{\sum_{i=1}^ {n} w_i(x;h) Y_i}{\sum_{i=1}^{n} w_i(x;h)} \]
where $w_i(x;h) = K_h(x - X_i) = \frac{1}{h}K\left(\frac{x - X_i}{h}\right)$, $K: \mathbb{R}^d \to \mathbb{R}$ is a kernel function, and $h = h(n) \in \mathbb{R}^+$ is the bandwidth.

\subsection{Estimation with Fixed Waypoints}

We consider an internal constraint model on the index set $I=\{1,2,\ldots,n\}$. We partition this set into two subsets: a set of deterministic fixed waypoints $C\subset I$ with cardinality $|C|=m$, and a complement set of stochastic observations $D=I\setminus C$.
For indices $i\in D$, the observations follow the standard stochastic nonparametric regression model: \begin{equation*} Y_i = m(X_i) + \varepsilon_i, \quad \mathbb{E}[\varepsilon_i] = 0, \end{equation*} where $m(x)=\mathbb{E}[Y|X=x]$ is the target regression function. Conversely, for indices $i \in C$, representing a set of fixed waypoints, the corresponding observations are assumed to be deterministic and lie exactly on the regression curve, i.e., $Y_i = m(X_i)$.

\subsection{Naive Bandwidth-Shrinking Strategy}

A natural baseline for incorporating fixed waypoints is to modify the local
smoothing scale. The intuition is to shrink the kernel bandwidth in the
neighborhood of fixed waypoints so as to force the estimator to adhere more
closely to the prescribed values.

Let $h>0$ be a global reference bandwidth and let
$\gamma:(0,\infty)\to(0,1]$ be a shrinkage factor.
For each evaluation location $x$ and observation $X_i$, we define a
point--dependent bandwidth
\[
h_i(x)=
\begin{cases}
\gamma(h), 
& \text{if } |x-X_i|\le r_h \text{ for some } i\in C,\\[4pt]
h, 
& \text{otherwise},
\end{cases}
\]

where $C$ denotes the index set of fixed waypoints and $r_h>0$ is a prescribed
neighborhood radius.

The resulting naive kernel weights are given by
\[
\tilde w_i(x;h)
=
\frac{1}{h_i(x)}
K\!\left(\frac{x-X_i}{h_i(x)}\right),
\]
leading to the estimator
\begin{equation}
\hat m_{\mathrm{naive}}(x)=
\frac{\sum_{i\in I}\tilde w_i(x;h)\,Y_i}
     {\sum_{i\in I}\tilde w_i(x;h)}.
\label{eq:naive}
\end{equation}

While intuitive, this approach tightly couples waypoint adherence with the
local smoothing scale. Shrinking $h_i(x)$ reduces the effective averaging
neighborhood around the evaluation location, which can substantially increase
local variance and lead to jagged artifacts near constrained regions.
Furthermore, selecting an appropriate shrinkage factor $\gamma(\cdot)$ is
often computationally expensive and lacks a robust theoretical guideline.

\subsection{Adaptive Nadaraya-Watson with fixed Waypoint tuning parameters (ANW)}

To overcome these limitations, we propose the Adaptive Nadaraya-Watson (ANW) estimator. Instead of altering the smoothing scale, ANW preserves a global bandwidth $h$ and incorporates fixed waypoint influence solely through adaptive reweighting. We define: \begin{equation*} \bar{w}_i(x;h) = \begin{cases} w_i(x;h), & i \in D, \\ \lambda_i w_i(x;h), & i \in C, \end{cases} \end{equation*} where $ w_i(x;h) = \frac{1}{h}K\!\left(\frac{x - X_i}{h}\right)$  and $\lambda_i\ge1$ is a tuning parameter controlling the ``strength'' of the $i$-th fixed waypoint. In principle, the waypoint influence parameters $\{\lambda_i : i \in C\}$
may vary across fixed waypoints, allowing different levels of enforcement.
In this work, however, we focus on the practically relevant case where a
single global tuning parameter is used and write $\lambda_i \equiv \lambda$
for all $i \in C$. The ANW estimator is the solution to the weighted local constant least squares problem: \begin{equation} \hat{m}_{{\mathrm{ANW}}}(x) = \arg\min_{B \in \mathbb{R}} \left[ \sum_{i \in D} (Y_i - B)^2 w_i(x;h) + \sum_{i \in C} (Y_i - B)^2 \lambda w_i(x;h) \right], \end{equation} yielding the closed-form expression: \begin{equation}\label{eq:anw} \hat{m}_{\mathrm{ANW}}(x) = \frac{\sum_{i \in D} w_i(x;h) Y_i + \sum_{i \in C} \lambda w_i(x;h) Y_i}{\sum_{i \in D} w_i(x;h) + \sum_{i \in C} \lambda w_i(x;h)}. \end{equation}

The primary advantage of ANW is the decoupling of smoothness and constraint adherence. As $\lambda \rightarrow\infty$, the estimator is forced through the fixed waypoint without shrinking the local window, thereby maintaining the global smoothness governed by $h$. This separation makes the ANW framework highly suitable for iterative data sharpening procedures for further bias reduction \cite{chen2025iterated,bs2025,shi2026}.

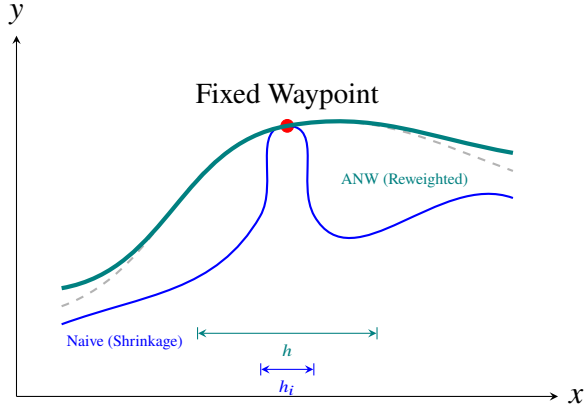
\begin{figure}[htbp]
    \centering
    \begin{tikzpicture}[scale=1.2, >=stealth]
        \draw[->] (0,0) -- (6,0) node[right] {$x$};
        \draw[->] (0,0) -- (0,4) node[above] {$y$};
        
        \draw[thick, dashed, gray!60] (0.5,1) to[out=20, in=190] (3,3) to[out=10, in=160] (5.5,2.5);

        \filldraw[red] (3,3) circle (2pt) node[above, black, yshift=2pt] {\small Fixed Waypoint};
        
        \draw[thick, blue] (0.5,0.8) to[out=20, in=240] (2.7,2) to[out=60, in=180] (3,3) to[out=0, in=120] (3.3,2) to[out=300, in=160] (5.5,2.2);

        \draw[ultra thick, teal] (0.5,1.2) to[out=10, in=190] (3,3) to[out=10, in=170] (5.5,2.7);
        
        \node[blue, font=\tiny] at (1.2,0.6) {Naive (Shrinkage)};
        \node[teal, font=\tiny] at (4.3,2.4) {ANW (Reweighted)};
        
        \draw[|<->|, blue] (2.7,0.3) -- (3.3,0.3) node[midway, below, font=\tiny] {$h_i$};
        \draw[|<->|, teal] (2,0.7) -- (4,0.7) node[midway, below, font=\tiny] {$h$};
    \end{tikzpicture}
    \caption{Comparison of fixed waypoint enforcement: Naive bandwidth shrinkage vs. ANW reweighting}
    \label{fig:anw_vs_naive}
\end{figure}

Fig.~\ref{fig:anw_vs_naive} provides a conceptual comparison between the two strategies for fixed waypoint enforcement. The naive strategy (blue line) reduces the bandwidth locally at the fixed waypoint, which forces the curve toward the constraint but results in a ``peaking'' artifact and a loss of smoothness. In contrast, the ANW estimator (green line) maintains a consistent global smoothing scale $h$ across the entire domain. By adjusting only the weight tuning parameter $\lambda$, the ANW estimator pulls the regression curve smoothly through the fixed waypoint without introducing local irregularities or coupling the constraint adherence to a reduction in the averaging neighborhood.

\subsection{Tuning of $(h,\lambda)$ on a standardized scale}

We use cross--validation to choose the optimal bandwidth for each method
\cite{stone1974cross}. To make the cross--validation error and the fixed
waypoint deviation comparable across different data sets, we standardize
the response to have zero mean and unit variance before fitting.
Specifically, let
\[
Y_i^\star = \frac{Y_i - \bar Y}{s_Y}, 
\qquad 
\bar Y = \frac{1}{n}\sum_{i=1}^n Y_i, 
\qquad 
s_Y^2 = \frac{1}{n}\sum_{i=1}^n (Y_i - \bar Y)^2,
\]
and let $\hat m^\star_{\mathrm{ANW},\lambda,h}(x)$ denote the ANW estimator
\eqref{eq:anw} constructed from the standardized responses
$\{(X_i, Y_i^\star)\}_{i=1}^n$. 

We then select $(\lambda,h)$ by minimizing
\begin{equation}
\mathrm{Loss}(\lambda, h)
=
\mathrm{CV}^\star(\lambda, h)
+
\sum_{i\in C}
\left(
\hat m^\star_{\mathrm{ANW},\lambda,h}(X_i)
-
Y_i^\star
\right)^2,
\label{eq:loss}
\end{equation}
where $\mathrm{CV}^\star(\lambda,h)$ denotes the $K$--fold cross--validation
error computed on the standardized scale. The standardized formulation implicitly
assigns equal weight to the cross--validation error and the waypoint
deviation. In practice, the relative importance of waypoint adherence
versus global smoothness can be adjusted by introducing an explicit
weighting parameter. We fix the weight to unity here to avoid introducing
an additional tuning parameter and to focus on the behavior of
$(\lambda,h)$.

\begin{figure}[htbp]
    \centering
    \includegraphics[width=\linewidth, trim=0cm 0cm 0cm 0.9cm, clip]{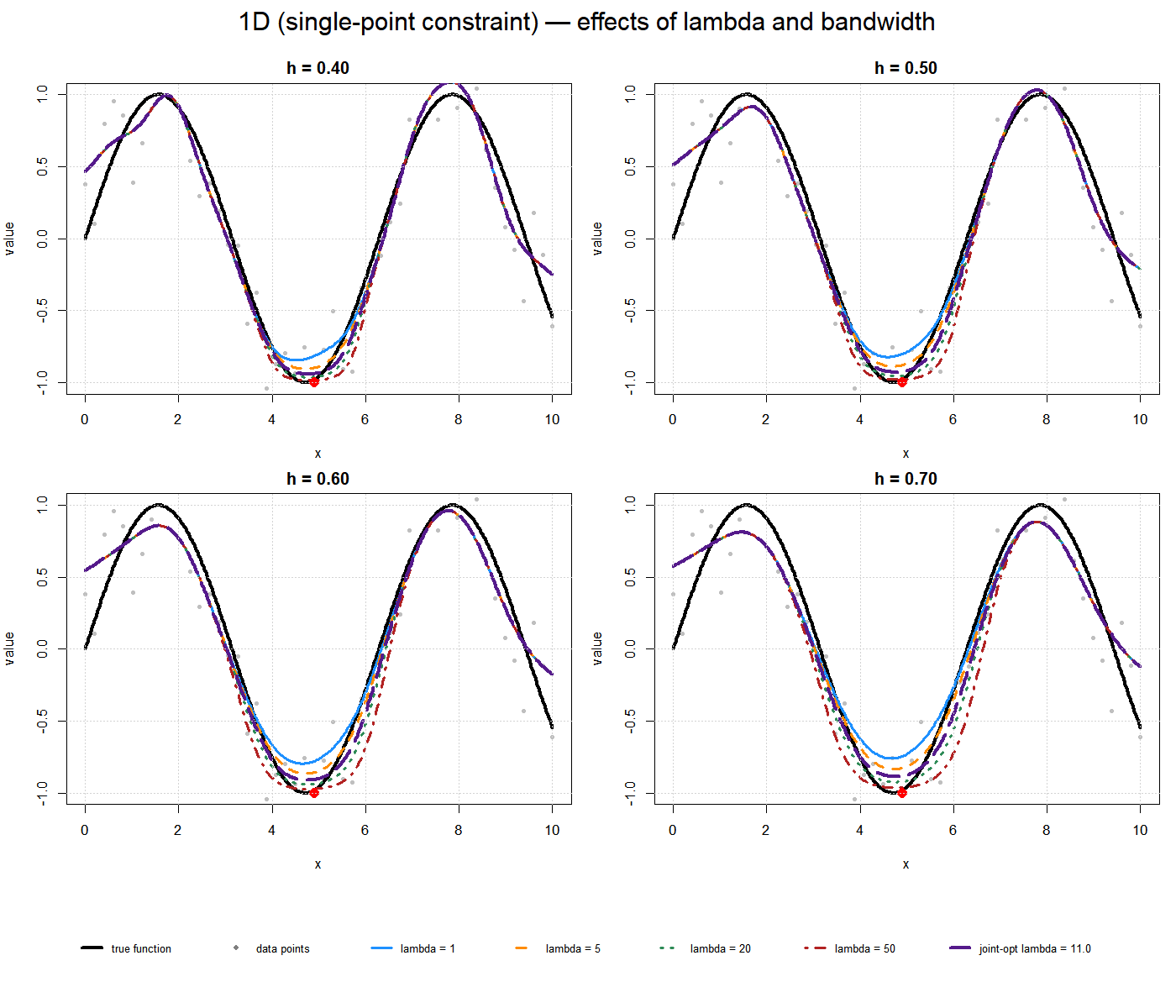}
    \caption{Effect of the penalty parameter $\lambda$ on the fitted curve under a single waypoint constraint}
    \label{fig:single}
\end{figure}

To illustrate the influence of the tuning parameter $\lambda$, we present a representative example in Fig.~\ref{fig:single}. As $\lambda$ increases, the fitted curve is progressively pulled toward the fixed waypoint (red circle), resulting in increasingly strict adherence to the prescribed value.

\section{Data sharpening for ANW (DS--ANW)}
\label{sec:dsanw}

\subsection{Motivation: bias reduction without changing weights}

The ANW estimator in \eqref{eq:anw} is a local constant kernel regression
(with fixed waypoint tuning parameters) and therefore inherits the familiar smoothing bias.
In particular, even when waypoint adherence is controlled by $\lambda$,
the leading bias term at interior points remains of order $O(h^2)$, as in
standard kernel regression. This motivates a bias--reduction device that does not change the weighting scheme $\bar w_i(x;h)$ and hence preserves
the fixed waypoint mechanism.

Data sharpening provides exactly such a strategy: rather than modifying the
estimator form or altering the bandwidth, it iteratively updates the responses
$\{Y_i\}^{n}_{i=1}$ using fitted residuals and then re-applies the same smoothing operator.
We adopt the iterated data sharpening procedure IDS2, which is designed to
reduce successive orders of the bias expansion while keeping variance inflation
modest \cite{chen2025iterated,bs2025,shi2026}. 

\subsection{IDS2 update rule under ANW weights}

Let $\hat m^{(0)}_{\mathrm{ANW}}(x)$ denote the original ANW estimator \eqref{eq:anw}
constructed from the observed responses $\{Y_i\}_{i=1}^n$.
Define the sharpened responses by $Y_i^{(0)} = Y_i$ and, for
$k=0,1,\ldots,M-1$, update
\begin{equation}
Y_i^{(k+1)}
=
Y_i
+
\left\{
Y_i^{(k)} - \hat m_{\mathrm{ANW}}^{(k)}(X_i)
\right\},
\qquad i=1,\ldots,n,
\label{eq:ids2_update}
\end{equation}

where $\hat m_{\mathrm{ANW}}^{(k)}(\cdot)$ denotes the ANW estimator applied to the
sharpened dataset $\{(X_i, Y_i^{(k)})\}_{i=1}^n$, with the fitted values
$\hat m_{\mathrm{ANW}}^{(k)}(X_i)$ obtained by evaluating the same estimator at
$x = X_i$ using the unchanged design points $\{X_i\}$ throughout. The defining feature of IDS2 is that the residual from
iteration $k$ is added back to the original response $Y_i$ rather than to
$Y_i^{(k)}$, which is crucial for bias reduction across iterations. 
Given $\{Y_i^{(k)}\}$, we compute
\begin{equation}
\hat m^{(k)}_{\mathrm{ANW}}(x)
=
\frac{\sum_{i\in I}\bar w_i(x;h)\,Y_i^{(k)}}
     {\sum_{i\in I}\bar w_i(x;h)},
\label{eq:dsanw_est}
\end{equation}
with the same adaptive weights
\[
\bar w_i(x;h)=
\begin{cases}
w_i(x;h), & i\in D,\\[3pt]
\lambda \, w_i(x;h), & i\in C,
\end{cases}
\qquad \lambda \ge 1.
\]
After $M$ iterations, the final estimator
\[
\hat m_{\mathrm{DS-ANW}}(x):=\hat m^{(M)}_{\mathrm{ANW}}(x)
\]
is referred to as the data--sharpened ANW (DS--ANW).

\subsection{Algorithm}

Equation \eqref{eq:ids2_update} has a simple interpretation. At iteration $k$,
the fitted value $\hat m^{(k)}_{\mathrm{ANW}}(X_i)$ represents the amount of smoothing
imposed by the ANW operator at $X_i$. The residual term
$Y_i^{(k)}-\hat m^{(k)}_{\mathrm{ANW}}(X_i)$ estimates what has been smoothed away.
IDS2 then adds this residual back to the original response $Y_i$, producing a
sharpened response that partially restores local curvature while keeping the
same weighting mechanism.

A key advantage in our setting is that sharpening does not alter the fixed waypoint
weights $\lambda$ nor the bandwidth $h$ inside $\bar w_i(x;h)$; hence the
fixed waypoint influence mechanism is preserved. Moreover, IDS2 was developed to
achieve systematic bias reduction across iterations, whereas the earlier IDS1
iteration (which adds residuals to the current sharpened responses) can fail to
improve bias beyond the first step. 

We also note a standard caveat: for local constant regression, the boundary
region grows with the number of sharpening steps, and bias improvement near
boundaries may be limited. In our applications, this
issue is mitigated by focusing inference and comparison on interior regions or
by using sufficiently dense sampling near endpoints.

\begin{algorithm}[htbp]
\caption{Data--Sharpened ANW (DS--ANW) via IDS2}
\label{alg:dsanw}
\begin{algorithmic}[1]
\STATE \textbf{Input:} data $\{(X_i,Y_i)\}_{i=1}^n$, fixed waypoint set $C$ (with $D=I\setminus C$), tuning parameter $\lambda$, bandwidth $h$, iterations $M$.
\STATE Initialize $Y_i^{(0)} \leftarrow Y_i$ for $i=1,\ldots,n$.
\FOR{$k = 0,1,\ldots,M-1$}
    \STATE Compute $\hat m^{(k)}_{\mathrm{ANW}}(x)$ from \eqref{eq:dsanw_est} using $\{(X_i,Y_i^{(k)})\}$ and weights $\bar w_i(x;h)$.
    \FOR{$i=1,\ldots,n$}
        \STATE Update $Y_i^{(k+1)} \leftarrow Y_i + \left(Y_i^{(k)} - \hat m^{(k)}_{\mathrm{ANW}}(X_i)\right)$.
    \ENDFOR
\ENDFOR
\STATE \textbf{Output:} $\hat m_{\mathrm{DS-ANW}}(x) \leftarrow \hat m^{(M)}_{\mathrm{ANW}}(x)$.
\end{algorithmic}
\end{algorithm}

\subsection{Simulation: DS--ANW improves ANW via bias reduction}
\label{sec:sim_dsanw}

In practice, $M$ is kept small (e.g., $M=1,2,3$) to balance bias reduction
against gradual variance inflation. The IDS2 study reports that additional
iterations can reduce estimation error and may also reduce sensitivity to
bandwidth choice in finite samples \cite{chen2025iterated}.
In our experiments, we treat $M$ as a small tuning parameter and compare
DS--ANW with the unsharpened ANW baseline.

We consider a one-dimensional alignment prototype under the regression model
\[
Y_i = m(X_i) + \varepsilon_i,\qquad
\varepsilon_i \stackrel{iid}{\sim} \mathcal{N}(0,\sigma^2),
\]
where $X_i \stackrel{iid}{\sim} U(0,1)$, $n=500$, and $\sigma=0.15$.
The ground-truth signal is
\[
m(x)=\sin(2\pi x)+0.3\cos(6\pi x).
\]
To mimic fixed waypoints, two observation locations are randomly selected.
Using the tuned parameters, we compare the baseline ANW estimator ($M=0$) with
its data--sharpened extensions DS--ANW obtained by applying IDS2 with
$M=1$ and $M=2$ iterations.

\begin{figure}[htbp]
    \centering
    \includegraphics[width=\linewidth, trim=0cm 0cm 0cm 0.9cm, clip]{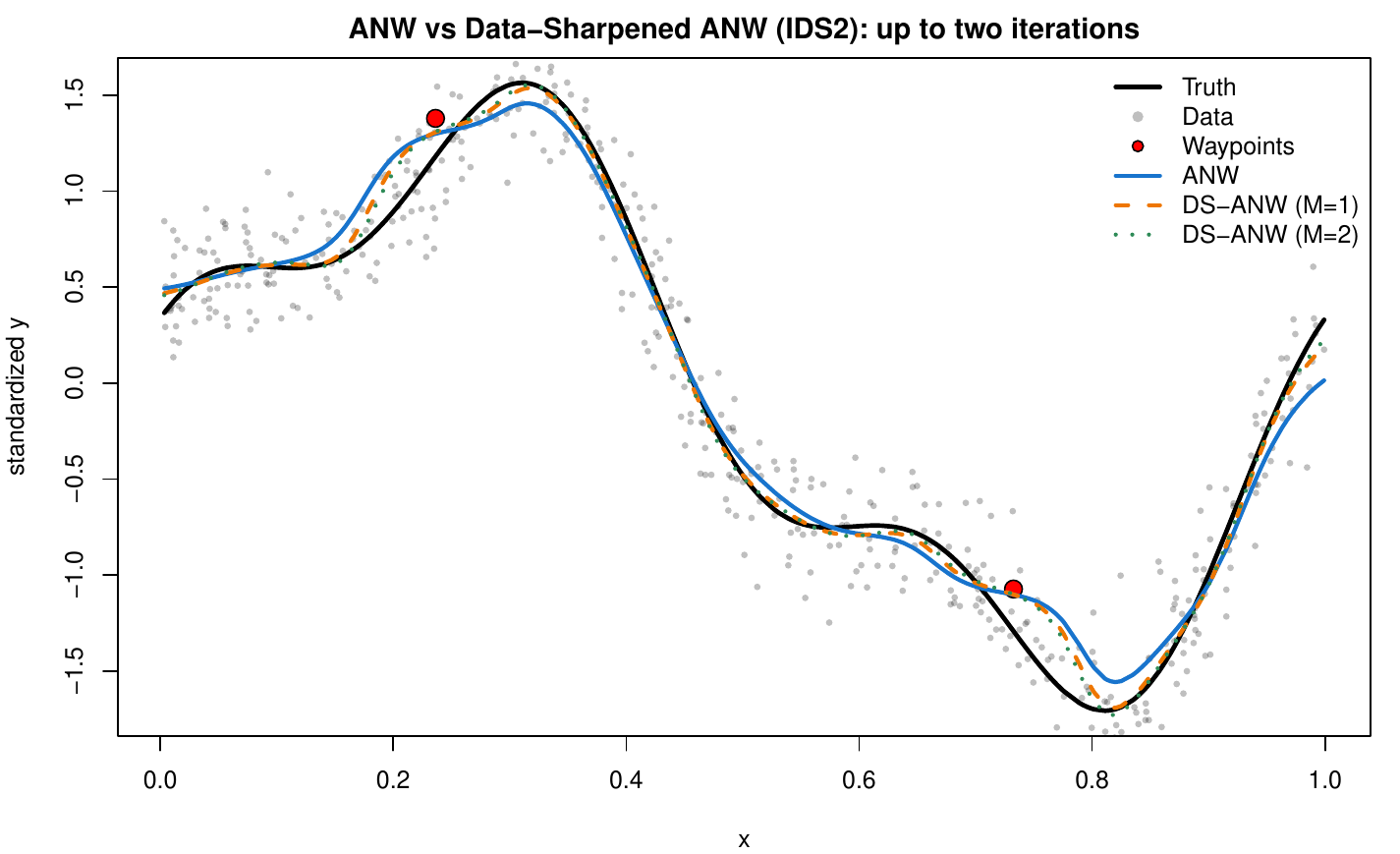}
    \caption{Comparison of ANW and DS-ANW via Iterative IDS2 Refinement}
    \label{fig:dsanw_curve}
\end{figure}
Fig.~\ref{fig:dsanw_curve} illustrates how successive sharpening steps reduce
smoothing bias while preserving the fixed waypoint attraction mechanism.

\begin{table}[htbp]
\centering
\caption{Performance comparison of ANW and DS--ANW with different sharpening iterations}
\label{tab:ds_anw_comparison}
\begin{tabular}{lccccc}
\hline
Method & $h$ & $\lambda$ & RMSE & WaypointError & Smoothness \\
\hline
ANW
& 0.030 & 100
& 0.127 & 0.080 & 0.245 \\

DS--ANW ($M=1$)
& 0.030 & 100
& 0.089 & 0.073 & 0.321 \\

DS--ANW ($M=2$)
& 0.030 & 100
& 0.080 & 0.069 & 0.335 \\

DS--ANW ($M=3$)
& 0.030 & 100
& 0.076 & 0.067 & 0.335 \\
\hline
\end{tabular}
\end{table}

Table~\ref{tab:ds_anw_comparison} summarizes the performance of ANW and DS--ANW under a fixed smoothing configuration, where the bandwidth $h$ and penalty parameter $\lambda$ are held constant across all sharpening iterations. The results illustrate the systematic effect of iterative data sharpening on global accuracy, fixed waypoint adherence, and smoothness.

As the number of sharpening iterations increases, DS--ANW achieves a substantial reduction in RMSE relative to ANW, indicating effective bias correction through successive updates of the response values. At the same time, the fixed waypoint error decreases monotonically from ANW to DS--ANW with $M=3$, demonstrating that moderate sharpening does not degrade, and may even slightly improve, adherence to the prescribed fixed waypoints under this configuration.

However, these gains in accuracy and fixed waypoint fidelity are accompanied by a noticeable loss of smoothness, with the curvature-based smoothness measure increasing sharply after the first sharpening step and stabilizing thereafter. In particular, the marginal improvements in RMSE and fixed waypoint error beyond $M=2$ become relatively small, while the smoothness penalty shows little further change.

Taken together, these results suggest that DS--ANW provides diminishing returns beyond a small number of sharpening iterations. In practice, one or two sharpening steps offer a favorable compromise between bias reduction and shape regularity, whereas additional iterations yield limited benefits at the cost of increased roughness.

\subsection{Asymptotic Properties}

We study the asymptotic behavior of the ANW estimator
$\hat B_{\mathrm{ANW}}(x)$ under standard regularity conditions
for kernel smoothing.

\begin{assumption}[Bandwidth]
\label{ass:bandwidth}
The bandwidth $h=h_n$ satisfies
\[
h \to 0
\quad \text{and} \quad
nh \to \infty
\qquad \text{as } n \to \infty .
\]
\end{assumption}

\begin{assumption}[Kernel]
\label{ass:kernel}
The kernel function $K:\mathbb{R}\to\mathbb{R}$ is symmetric,
satisfies
\[
\int K(u)\,du = 1,
\qquad
\int u^2 K(u)\,du = \sigma_K^2 < \infty,
\qquad
\int K^2(u)\,du = R(K) < \infty .
\]
\end{assumption}

\begin{assumption}[Smoothness]
\label{ass:smoothness}
The regression function $m(x)=\mathbb{E}[Y\mid X=x]$
and the marginal density $f(x)$ of $X$
are twice continuously differentiable in a neighborhood of $x$,
with $f(x)>0$.
\end{assumption}

\begin{assumption}[Errors]
$\varepsilon=Y-m(X)$ has conditional variance $\sigma^2(x)=\mathrm{Var}(Y\mid X=x)$ continuous and bounded near $x$.
\end{assumption}

\begin{assumption}[Fixed Waypoint tuning parameter scaling]
To avoid asymptotic domination at interior points \cite{zhou2024adaptive},
\[
\lambda\to\infty
\quad\text{and}\quad
\frac{\lambda}{nh^4}\to 0
\qquad (n\to\infty).
\]
\end{assumption}

\begin{assumption}[Number of Constraints]
The number of fixed waypoints $q$ is fixed and does not grow with $n$.
\end{assumption}

A central requirement of fixed--waypoint regression is that the estimator
must recover the prescribed values exactly at the constraint locations,
at least asymptotically. 
While the adaptive Nadaraya--Watson (ANW) estimator incorporates fixed
waypoints through adaptive reweighting, it is not a priori obvious that
this mechanism guarantees convergence to the imposed values as the sample
size increases.

The following theorem establishes that, under standard kernel regularity
conditions and mild assumptions on the constraint weights, the ANW estimator
is pointwise consistent at fixed waypoint locations.

\begin{theorem}[Pointwise convergence at fixed waypoints]
\label{thm:waypoint_convergence}
Under Assumptions~1,~2,~5, and~6,
\begin{equation}
 \hat m_{\mathrm{ANW}}(X_j) \xrightarrow{p} Y_j,
\qquad j \in C .   
\end{equation}

\end{theorem}

Theorem~\ref{thm:waypoint_convergence} shows that the adaptive weighting
mechanism of ANW enforces the fixed waypoint constraints in a strong
pointwise sense. 
In particular, the estimator does not merely remain close to the waypoint
values, but converges in probability to the imposed response exactly at
each constrained location.

This result provides a theoretical justification for using ANW as a
constraint--preserving smoother in alignment design problems, where
exact adherence to prescribed waypoints is required.

While Theorem~\ref{thm:waypoint_convergence} characterizes the behavior of the
ANW estimator at fixed waypoint locations, it is equally important to
understand its asymptotic properties away from the constraints.
In particular, one may ask whether the adaptive reweighting mechanism
introduced to enforce fixed waypoints alters the classical interior
bias--variance trade--off of kernel regression.

The following theorem shows that, at interior points not subject to
constraints, the ANW estimator retains the same first--order asymptotic
behavior as the standard Nadaraya--Watson estimator.

\begin{theorem}[Interior asymptotics]
\label{thm:interior_asymptotics}
Under Assumptions~1--5, at interior points
$x \notin \{X_j : j \in C\}$, the ANW estimator satisfies
\begin{equation}
\label{eq:anw_bias}
\mathrm{Bias}\{\hat B_{\mathrm{ANW}}(x)\}
=
\frac{1}{2} h^2 \sigma_K^2
\left(
m''(x) + \frac{2 m'(x) f'(x)}{f(x)}
\right)
+ o(h^2),
\end{equation}
\begin{equation}
\label{eq:anw_var}
\mathrm{Var}\{\hat B_{\mathrm{ANW}}(x)\}
=
\frac{\sigma^2(x) R(K)}{n h f(x)}
+ o\!\left(\frac{1}{n h}\right),
\end{equation}
and
\begin{equation}
\label{eq:anw_clt}
\sqrt{n h}
\left(
\hat B_{\mathrm{ANW}}(x)
- m(x)
- \mathrm{Bias}\{\hat B_{\mathrm{ANW}}(x)\}
\right)
\xrightarrow{d}
N\!\left(
0,
\frac{\sigma^2(x) R(K)}{f(x)}
\right).
\end{equation}
\end{theorem}

Consequently, $\hat B_{\mathrm{ANW}}$ is asymptotically equivalent to the
Nadaraya--Watson estimator $\hat m_{\mathrm{NW}}$ at interior points
\cite{fan1996local}. In particular, with the usual choice
$h_{\mathrm{opt}}\asymp n^{-1/5}$, we have $nh^4\asymp n^{1/5}$, so that
$\lambda \to \infty$ while $\lambda/(nh^4)\to 0$ can hold simultaneously.
Hence, the enforcement of fixed waypoints does not inflate the interior
bias or variance asymptotically, and the effect of the constraints remains
localized to neighborhoods of the prescribed waypoints.

The complete mathematical proofs for Theorem \ref{thm:waypoint_convergence} and Theorem \ref{thm:interior_asymptotics} are provided in {Appendix}.

\section{Numerical Properties}

\subsection{One-dimensional fixed-waypoint regression (1D)}

First, we conducted a simple scenario validation to assess the performance of
ANW and DS--ANW in balancing fixed waypoint proximity, smoothness, and overall
estimation error.

The experimental data were generated from a known test function exhibiting
both oscillatory behavior and distinct structural features:
\begin{equation}
Y_i = m(X_i) + \varepsilon_i, \quad
m(x) = \sin x + 0.3\cos(2x), \quad x \in [0,10],
\end{equation}
where the error terms are independent and identically distributed,
$\varepsilon_i \stackrel{iid}{\sim} \mathcal{N}(0,\sigma^2)$.

 The sample size was set to 500, a moderately large dataset. Two cases with small and large waypoint deviations, respectively, were considered to demonstrate the ability of ANW and DS--ANW to approach fixed waypoints. We compared the standard Nadaraya–Watson kernel regression (NW), its adaptive constrained extension (ANW), and the data–sharpened variant (DS–ANW), together with local linear kernel regression (LLR) and spline-based methods. We evaluate the performance of the estimator using four metrics. The root mean squared error (RMSE) is a commonly used global fit accuracy metric. To evaluate the ANW estimator's ability to satisfy point value constraints, we computed the error at the constrained points (Waypoint Error). Additionally, the smoothness of the fitted curve is measured based on a second-order derivative penalty that quantifies curvature \cite{GreenSilverman1993}.

We first consider a mild constraint scenario (Case~1), where the two fixed
waypoints are selected such that their residuals relative to an unconstrained
baseline fit are small.
In this setting, the imposed waypoint constraints are close to the natural
data trend and therefore do not strongly conflict with global smoothness.

\begin{figure}[htbp]
    \centering
    \includegraphics[width=\linewidth, trim=0cm 0cm 0cm 0.9cm, clip]{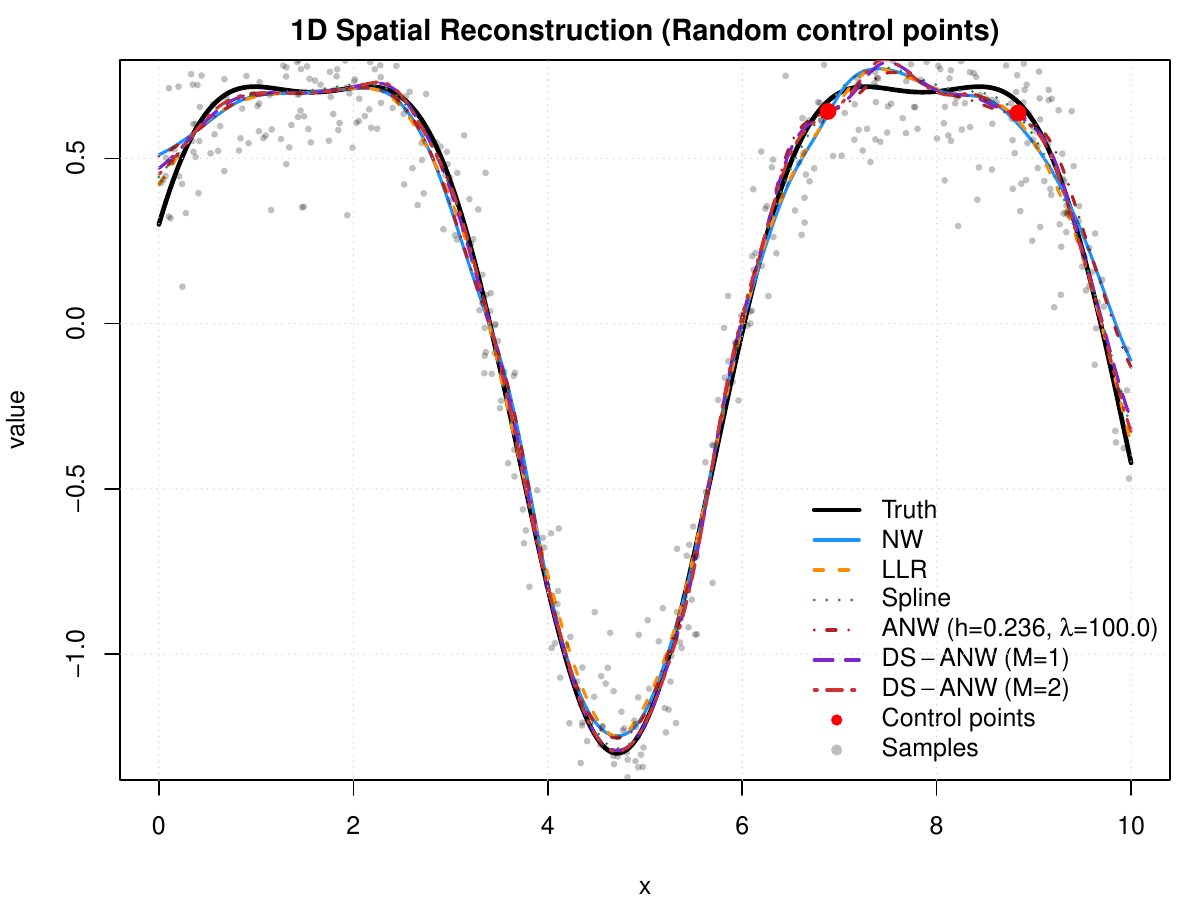}
    \caption{One-dimensional constrained regression with two fixed waypoints (Case~1)}
    \label{fig:1D1}
\end{figure}

As shown in Fig.~\ref{fig:1D1}, when the imposed waypoint constraints are mild,
most smoothing methods produce visually similar fits and achieve comparable
global accuracy.
In this regime, the fixed waypoints do not strongly conflict with the overall
data structure, and differences among constrained and unconstrained methods
are relatively small.

\begin{figure}[htbp]
    \centering
    \includegraphics[width=\linewidth, trim=0cm 0cm 0cm 0.9cm, clip]{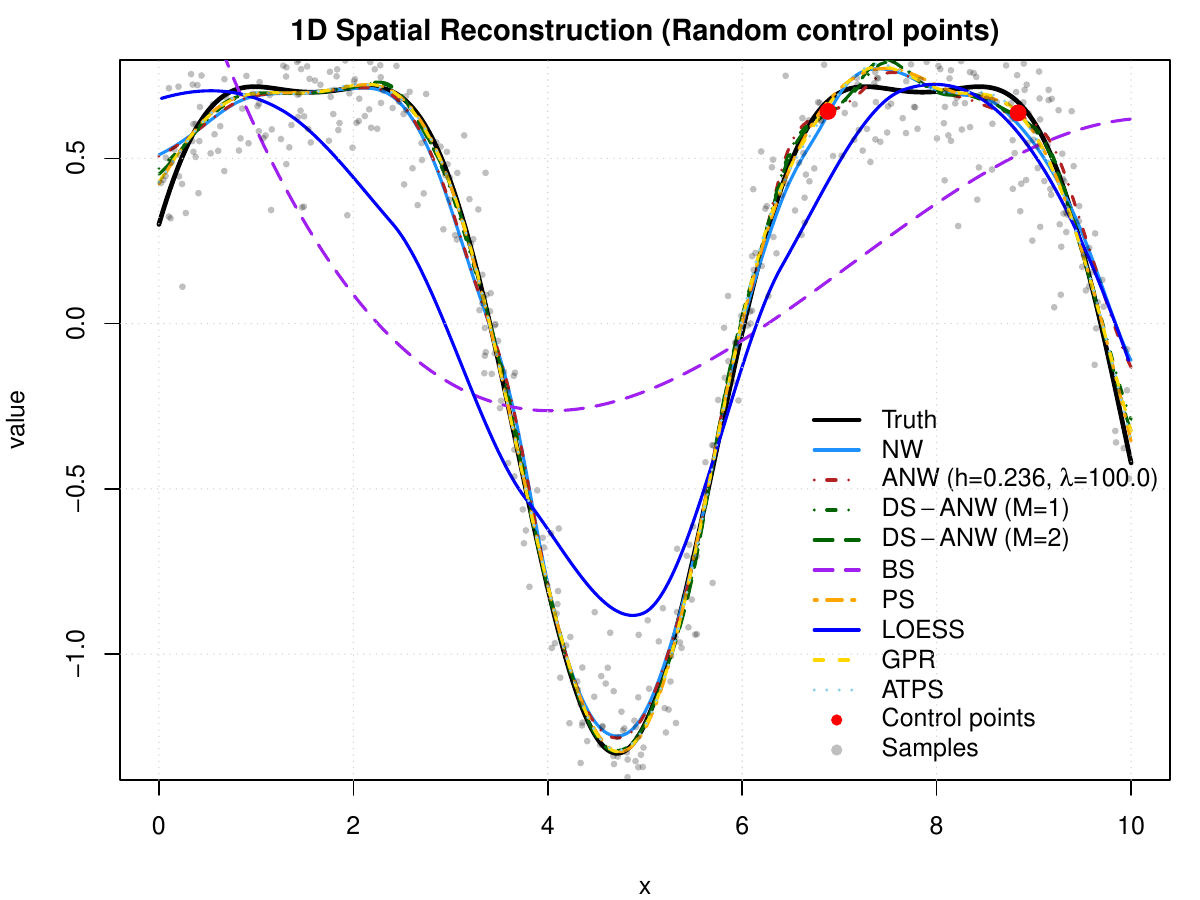}
    \caption{Comparison of one-dimensional nonparametric smoothing methods under fixed waypoint constraints (Case~1).}

    \label{fig:1D1_case1_com}
\end{figure}

We compare six representative nonparametric smoothing methods with ANW and DS-ANW in Fig.~\ref{fig:1D1_case1_com}. 
B-spline regression~\cite{de1977package} and smoothing splines~\cite{qiu1995nonparametric} are included as classical global smoothers, while P-splines~\cite{pya2015shape} represent a modern penalized spline framework.
LOESS~\cite{cleveland1988locally} serves as a canonical local regression method, Gaussian process regression (GPR)~\cite{seeger2004gaussian} represents a probabilistic kernel-based approach, and adaptive thin-plate splines (TPS)~\cite{wood2003thin} provide an adaptive spatial smoothing baseline.

To jointly evaluate fixed waypoint adherence, global accuracy, and smoothness,
we define a composite score based on three metrics:
\begin{equation}
\label{eq:CSS}
\mathrm{CSS}
= w_c \, \phi\!\left( \frac{\mathrm{Waypoint\ Error}}{\tau_c} \right)
+ w_g \, \phi\!\left( \frac{\mathrm{RMSE}}{\tau_g} \right)
+ w_s \, \phi\!\left( \frac{\mathrm{Smoothness}}{\tau_s} \right),
\end{equation}
where the penalty function $\phi(\cdot)$ is given by
\begin{equation}
\label{eq:phi}
\phi(x) = \max(0,\, x - 1).
\end{equation}
Each tolerance parameter $\tau_c$, $\tau_g$, and $\tau_s$ represents an
acceptable upper bound for the corresponding metric, beyond which a linear
penalty is incurred. In our experiments, the waypoint error tolerance is set to
$\tau_c=0.10$, which is stricter than the noise standard deviation
($\sigma=0.15$) and thus reflects a demanding waypoint requirement under
measurement noise. The global error tolerance is set to $\tau_g=0.05$.
The smoothness tolerance $\tau_s$ is chosen as the median smoothness value
across all candidate fits in the experiment. The weights are fixed at
$w_c=0.5$, $w_g=0.3$, and $w_s=0.2$.

\begin{table}[htbp]
\centering
\caption{Performance comparison of estimators in 1D simulation (Case~1)}

\label{tab:1d_main_case1}
\begin{tabular}{lcccc}
\hline
Model 
& RMSE 
& Waypoint Error 
& Smoothness 
& CSS \\
\hline
NW     
& 0.060 
& 0.025 
& $4.43\times10^{-7}$ 
& 0.10 \\

ANW    
& 0.056 
& 0.006 
& $6.11\times10^{-7}$ 
& 0.17 \\

DS--ANW ($M=1$)
& 0.034
& 0.008
& $6.22\times10^{-7}$
& 0.14 \\

DS--ANW ($M=2$)
& 0.032
& 0.009
& $6.37\times10^{-7}$
& 0.15 \\

BS     
& 0.553 
& 0.725 
& $1.03\times10^{-8}$ 
& 6.14 \\

PS     
& 0.026 
& 0.021 
& $3.63\times10^{-7}$ 
& 0.00 \\

LOESS 
& 0.236 
& 0.156 
& $1.31\times10^{-7}$ 
& 1.39 \\

GPR   
& 0.027 
& 0.013 
& $3.66\times10^{-7}$ 
& 0.00 \\

ATPS  
& 0.028 
& 0.006 
& $3.86\times10^{-7}$ 
& 0.01 \\
\hline
\end{tabular}
\end{table}

In this case, the fixed waypoints are chosen such that their deviations from a baseline fit are small, corresponding to a relatively mild and easy constraint scenario. Consequently, most methods achieve low global errors and satisfy the fixed waypoints reasonably well.

As shown in Table~\ref{tab:1d_main_case1}, penalized and probabilistic smoothers, including P-splines, Gaussian process regression (GPR), and adaptive thin-plate splines (ATPS), attain very small RMSE values and near-zero composite scores (CSS), reflecting strong overall fits when the imposed constraints are weak. The standard NW estimator also performs competitively in this regime, with moderate constraint deviations.

Although performance differences across methods are less pronounced in this setting, ANW achieves the smallest waypoint error among the kernel-based estimators while maintaining comparable global accuracy. This indicates that adaptive weighting can further stabilize waypoint adherence even when the baseline alignment is already strong.

The data-sharpened ANW estimators (DS--ANW with $M=1$ and $M=2$) further reduce RMSE relative to both NW and ANW, demonstrating effective bias reduction through iterative sharpening. While DS--ANW introduces a slight increase in waypoint error compared to ANW, the deviations remain small and stable across iterations, and the overall smoothness is preserved. These results suggest that DS--ANW offers a favorable trade-off between global accuracy and constraint adherence in low-noise, weakly constrained scenarios.

We next consider a challenging constraint scenario (Case~2), in which the two
fixed waypoints are deliberately chosen to have large residuals relative to an
unconstrained baseline fit.
In this setting, the imposed waypoint requirements strongly conflict with the
global data trend, making strict constraint enforcement substantially more
difficult.

\begin{figure}[htbp]
    \centering
    \includegraphics[width=\linewidth, trim=0cm 0cm 0cm 0.9cm, clip]{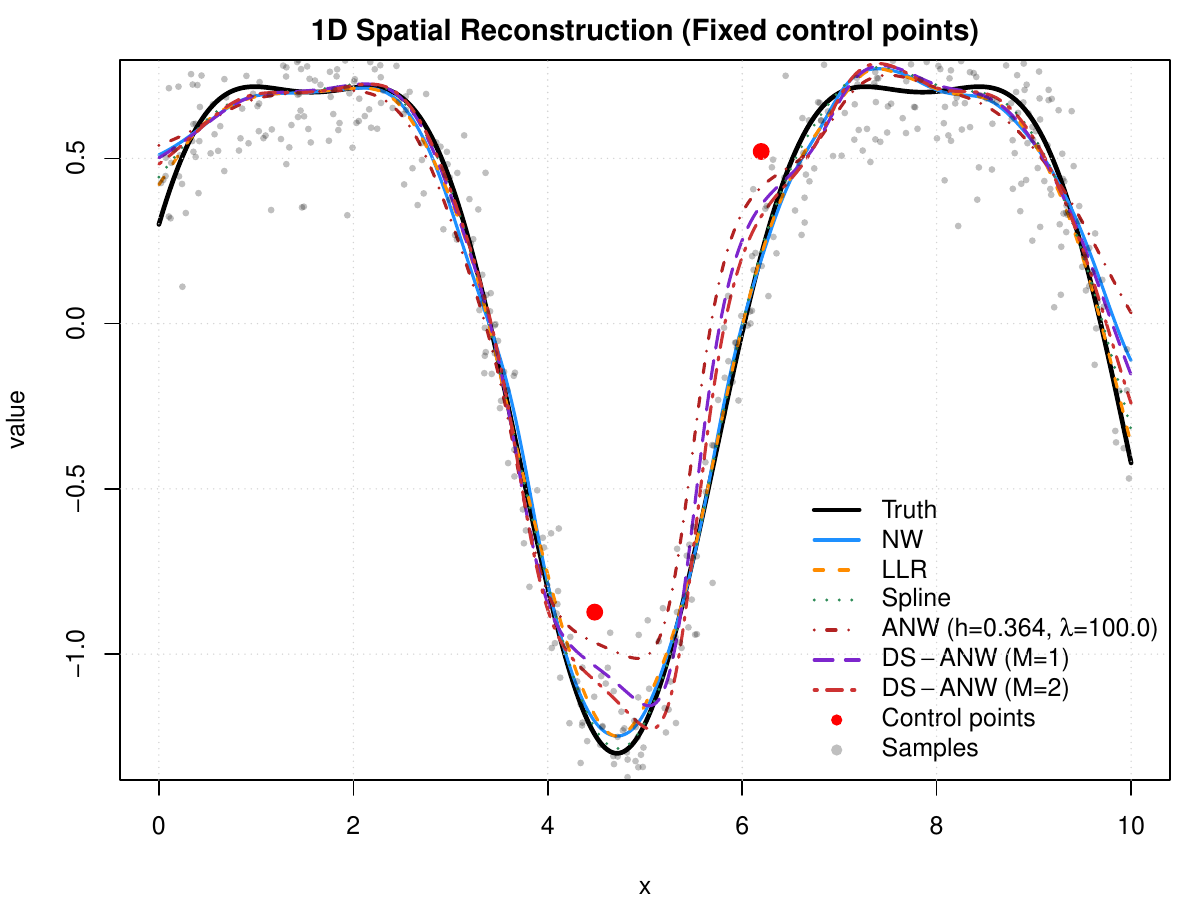}
    \caption{One-dimensional constrained regression with two fixed waypoints (Case~2)}
    \label{fig:1D2}
\end{figure}

As illustrated in Fig.~\ref{fig:1D2}, ANW is able to pull the fitted curve
substantially closer to the prescribed waypoints through adaptive reweighting,
even under strong conflict with the global structure.
In contrast, the data--sharpened variants (DS--ANW) progressively weaken
waypoint adherence as the number of sharpening iterations increases, reflecting
the trade-off between bias reduction and constraint enforcement in this regime.

\begin{figure}[htbp]
    \centering
    \includegraphics[width=\linewidth, trim=0cm 0cm 0cm 0.9cm, clip]{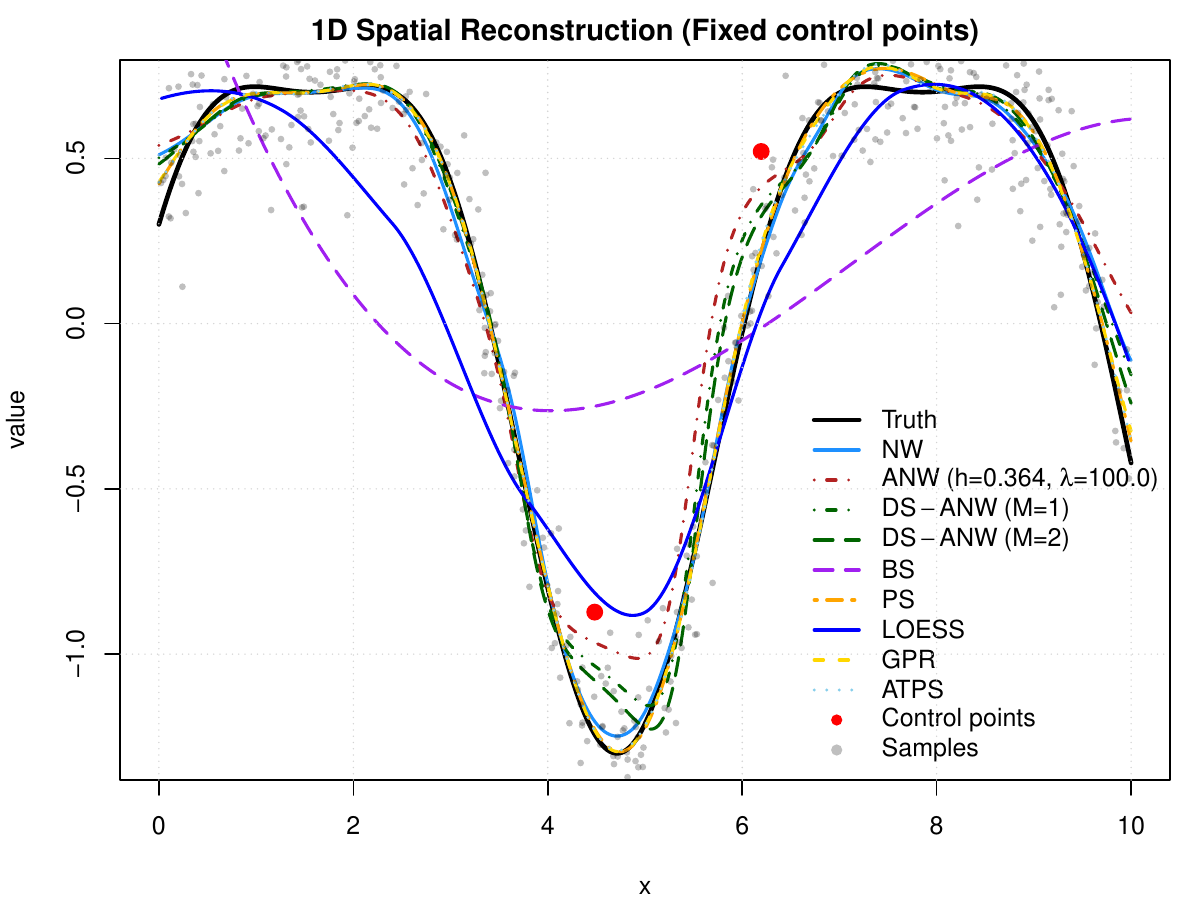}
    \caption{Comparison of one-dimensional nonparametric smoothing methods under fixed waypoint constraints (Case~2)}

    \label{fig:1D2_case2_com}
\end{figure}

Fig.~\ref{fig:1D2_case2_com} further compares ANW and DS--ANW with several
representative nonparametric smoothing methods.
Methods that do not explicitly incorporate waypoint information remain largely
insensitive to the imposed constraints and fail to approach the prescribed
locations, despite achieving low global RMSE in some cases.

\begin{table}[htbp]
\centering
\caption{Performance comparison of estimators in 1D simulation (Case~2)}

\label{tab:1d_main}
\begin{tabular}{lcccc}
\hline
Model 
& RMSE 
& Waypoint Error 
& Smoothness 
& CSS \\
\hline
NW     
& 0.060 
& 0.329 
& $4.43\times10^{-7}$ 
& 1.25 \\

ANW    
& 0.159 
& 0.100 
& $9.46\times10^{-7}$ 
& 0.97 \\

DS--ANW ($M=1$)
& 0.103
& 0.162
& $1.20\times10^{-6}$
& 1.08 \\

DS--ANW ($M=2$)
& 0.082
& 0.202
& $1.24\times10^{-6}$
& 1.18 \\

BS     
& 0.553 
& 1.691 
& $1.03\times10^{-8}$ 
& 10.97 \\

PS     
& 0.026 
& 0.328 
& $3.63\times10^{-7}$ 
& 1.14 \\

LOESS 
& 0.236 
& 0.345 
& $1.31\times10^{-7}$ 
& 2.34 \\

GPR   
& 0.027 
& 0.324 
& $3.66\times10^{-7}$ 
& 1.12 \\

ATPS  
& 0.028 
& 0.320 
& $3.86\times10^{-7}$ 
& 1.11 \\
\hline
\end{tabular}
\end{table}

As shown in Table~\ref{tab:1d_main}, ANW attains the lowest composite score (CSS) among all competing methods by achieving a favorable balance between global accuracy, constraint satisfaction, and smoothness. In this case, the fixed waypoints are deliberately selected to have large residuals relative to a baseline fit, representing a challenging constraint scenario in which pointwise requirements strongly conflict with global smoothness.

Under such conditions, most unconstrained or weakly constrained smoothers, including NW, P-splines, Gaussian process regression (GPR), LOESS, and adaptive thin-plate splines (ATPS), exhibit substantially increased waypoint errors despite maintaining low RMSE values. This indicates that global fit quality alone is insufficient to accommodate distant fixed waypoints. In contrast, ANW explicitly reweights the fixed waypoints through its adaptive penalty mechanism, leading to a pronounced reduction in waypoint error compared to all other methods and, consequently, the smallest CSS.

The data-sharpened ANW estimators (DS--ANW with $M=1$ and $M=2$) substantially reduce RMSE relative to ANW, reflecting effective bias correction through iterative sharpening. However, this improvement in global accuracy is accompanied by increased waypoint error and reduced smoothness as the number of sharpening iterations increases. As a result, DS--ANW yields larger CSS values than ANW in this strongly constrained setting, indicating that aggressive bias reduction may partially counteract the intended enforcement of distant fixed waypoints.

Overall, these results highlight a clear distinction between the roles of ANW and DS--ANW: ANW is better suited for scenarios dominated by strong or conflicting fixed waypoints, whereas DS--ANW provides a bias--variance trade-off that favors global accuracy but may compromise strict constraint adherence when the imposed fixed waypoints deviate substantially from the baseline structure.

\begin{table}[htbp]
\centering
\caption{Sensitivity analysis of ANW with respect to $q$ and $\lambda$ (Case~2)}
\label{tab:1d_sensitivity}
\begin{tabular}{cccc}
\hline
\textbf{$q$} & \textbf{$\lambda$} & \textbf{RMSE} & \textbf{Waypoint Error} \\
\hline
1 & 10   & 0.06 & 0.33 \\
1 & 100  & 0.10 & 0.10 \\
1 & 1000 & 0.14 & \textbf{0.01} \\
2 & 10   & 0.06 & 0.30 \\
2 & 100  & 0.12 & 0.09 \\
2 & 1000 & 0.17 & \textbf{0.01} \\
3 & 10   & 0.07 & 0.30 \\
3 & 100  & 0.14 & 0.09 \\
3 & 1000 & 0.20 & \textbf{0.01} \\
\hline
\end{tabular}
\end{table}

Table~\ref{tab:1d_sensitivity} reports a sensitivity analysis of ANW with respect
to the waypoint penalty parameter $\lambda$ and the number of fixed waypoints $q$
(Case~2). For each fixed $q$, increasing $\lambda$ strengthens constraint
enforcement, leading to a monotonic decrease in the waypoint error . This improved local
adherence is accompanied by an increase in global RMSE, indicating the expected
trade-off between fitting accuracy and exact waypoint satisfaction. Moreover,
for a fixed $\lambda$, using more waypoints (larger $q$) generally makes the
problem more restrictive and tends to increase RMSE, while the waypoint error
remains comparable across $q$ and is ultimately dominated by the choice of
$\lambda$.


\subsection{Two-dimensional constrained trajectory fitting (2D)}

To simulate spatial scenarios, we generate two-dimensional trajectories
$(x(s), y(s))$ parameterized by the path variable $s \in [0,1]$
\cite{mellinger2011minimum}.
A scalar response varying along the route is then modeled as
\[
z(s) = m(s) + \varepsilon(s), \qquad
\varepsilon(s) \stackrel{iid}{\sim} \mathcal{N}(0,\sigma^2).
\]

Although the data resides in two-dimensional space, regression analysis is still performed with $s$ as the variable, which naturally represents the position along the trajectory \cite{ramsay2005fda}. Under the condition of three fixed waypoints ($q=3$), we compared the standard NW estimator with ANW and DS--ANW. In the two-dimensional setting, bandwidth and penalty parameters are selected
separately for ANW and each DS--ANW variant via cross-validation, reflecting the
increased complexity of spatial trajectories.

\begin{figure}[htbp]
    \centering
    \includegraphics[width=0.5\linewidth, trim=0cm 0cm 0cm 0.9cm, clip]{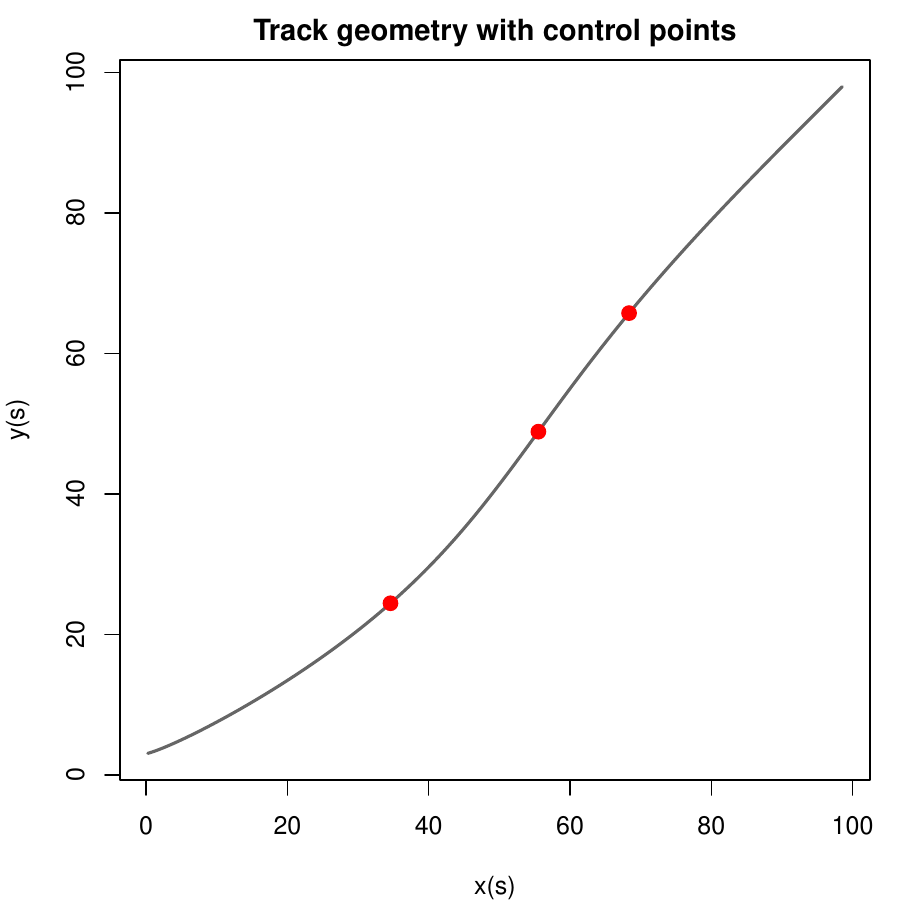}
            \caption{Track geometry with fixed waypoints}
    \label{fig:2d_track}
\end{figure}

\begin{figure}[htbp]
    \centering
    \includegraphics[width=\linewidth, trim=0cm 0cm 0cm 0.9cm, clip]{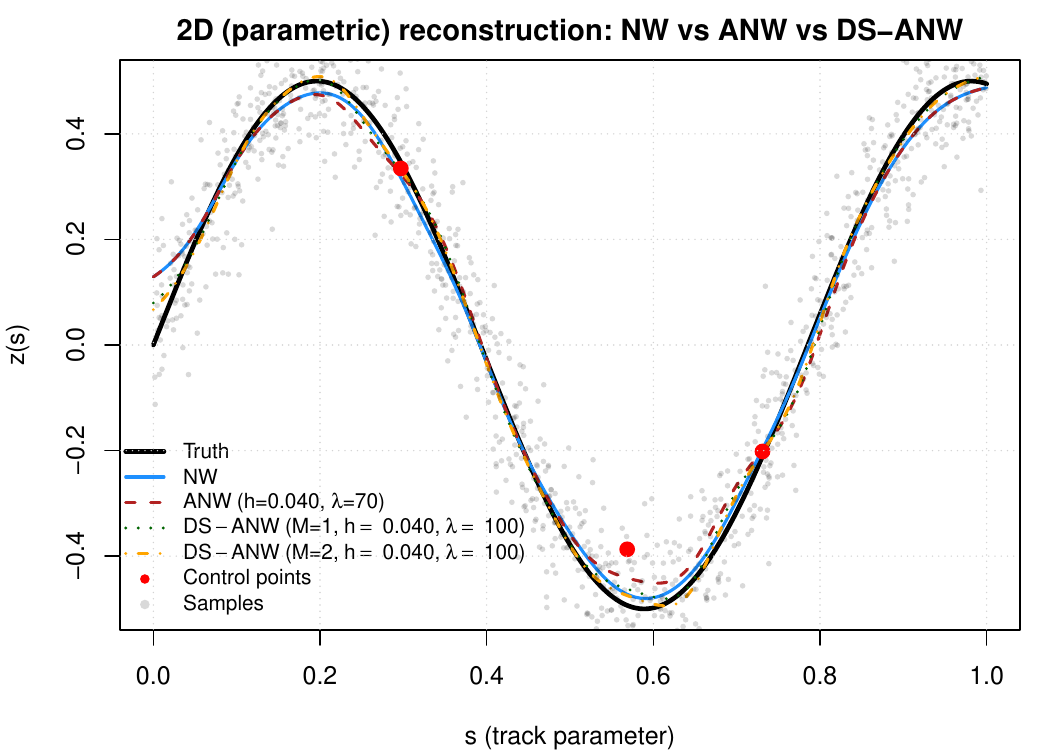}
    \caption{Two-dimensional constrained path fitting with fixed waypoints}
    \label{fig:2d_main}
\end{figure}

Fig.~\ref{fig:2d_track} illustrates a two-dimensional track geometry with several prescribed fixed waypoints imposed as spatial constraints. In Fig.~\ref{fig:2d_main}, the adaptive Nadaraya–Watson (ANW) mechanism is illustrated in a two-dimensional spatial setting, where the fitted surface is locally pulled toward the prescribed fixed waypoints to satisfy the imposed constraints. The data-sharpened variants (DS--ANW) further refine the fit through iterative bias reduction while preserving the same constraint-enforcement mechanism.

\begin{figure}[htbp]
    \centering
    \includegraphics[width=\linewidth, trim=0cm 0cm 0cm 0.9cm, clip]{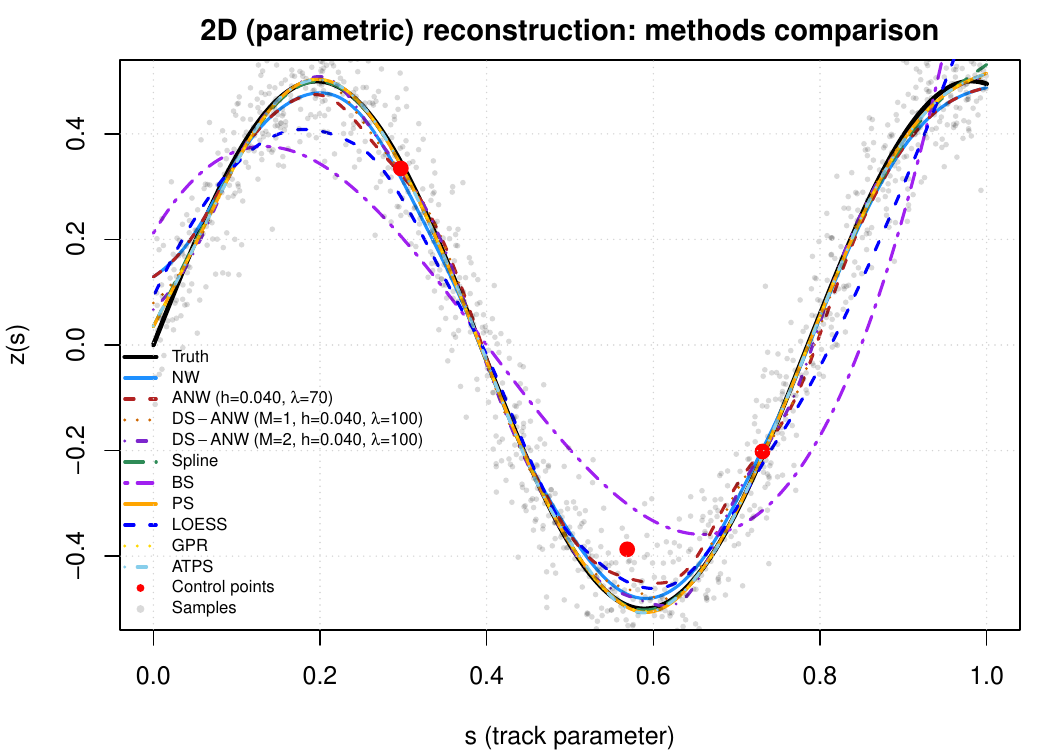}
    \caption{Comparison of two-dimensional constrained path fitting with fixed fixed waypoints}
    \label{fig:2d_main_com}
\end{figure}

Fig.~\ref{fig:2d_main_com} provides a comparative visualization of ANW against the alternative smoothing methods introduced earlier, highlighting their differing behaviors in the presence of fixed fixed waypoints.

\begin{table}[htbp]
\centering
\caption{Performance comparison of estimators in the 2D spatial simulation}
\label{tab:2d_main}
\begin{tabular}{lcccc}
\hline
Model & RMSE & Waypoint Error & Smoothness & CSS \\
\hline
NW     
& 0.02 
& 0.05 
& $8.19\times10^{-9}$ 
& 0.01 \\

ANW    
& 0.03 
& 0.03 
& $1.31\times10^{-8}$ 
& 0.14 \\

DS--ANW ($M=1$)
& 0.02
& 0.04
& $1.76\times10^{-8}$
& 0.25 \\

DS--ANW ($M=2$)
& 0.01
& 0.05
& $1.68\times10^{-8}$
& 0.23 \\

Spline 
& 0.01 
& 0.06 
& $7.10\times10^{-9}$ 
& 0.00 \\

BS     
& 0.16 
& 0.43 
& $5.55\times10^{-9}$ 
& 2.27 \\

PS     
& 0.01 
& 0.06 
& $7.56\times10^{-9}$ 
& 0.00 \\

LOESS  
& 0.07 
& 0.05 
& $6.89\times10^{-9}$ 
& 0.09 \\

GPR    
& 0.01 
& 0.06 
& $7.61\times10^{-9}$ 
& 0.00 \\

ATPS   
& 0.01 
& 0.07 
& $7.88\times10^{-9}$ 
& 0.00 \\
\hline
\end{tabular}
\end{table}

Table~\ref{tab:2d_main} summarizes the performance of representative estimators
in the 2D spatial simulation.
Global smoothers and probabilistic methods, including spline regression,
P-splines, GPR, and ATPS, achieve the smallest RMSE values, indicating strong
global fitting accuracy when assessed solely by prediction error.
However, these methods generally exhibit larger waypoint errors, suggesting
limited flexibility in accommodating externally imposed fixed waypoints.

ANW substantially reduces the waypoint error relative to unconstrained
smoothers, reflecting the effectiveness of its adaptive reweighting mechanism
in enforcing waypoint adherence.
This improvement is accompanied by an increase in RMSE compared with the best
global smoothers, illustrating the trade-off between global accuracy and
constraint satisfaction in the 2D setting.

The data-sharpened variants DS--ANW further reduce RMSE relative to ANW,
particularly for smaller sharpening iterations.
At the same time, increasing the sharpening level leads to a partial loss of
waypoint accuracy, highlighting the bias--variance--constraint trade-off
introduced by data sharpening.
Overall, ANW and DS--ANW provide a controlled compromise between global fit and
waypoint enforcement, yielding competitive composite scores when precise
constraint adherence is of practical importance.

\section{Empirical Applications}

In the route--alignment application, externally imposed fixed waypoints
(e.g., the requirement that the railway pass through Banff)
are incorporated into our framework by treating them as deterministic
fixed waypoints. Although these locations do not originate from the
observed trajectory data, they are augmented into the observation set
and included in the constraint set $C$.
By assigning them large penalty weights $\lambda$,
the estimated curve is effectively forced to interpolate these
prescribed locations. Importantly, because the augmented points are
treated as ordinary observations, the same constraint information is
made available to all competing methods, ensuring a fair and consistent
comparison. Under this formulation, external geometric requirements are
naturally handled as internal fixed waypoints within the proposed
model.

\subsection{Railway alignment design}

In the applications, the available data consist only of geographic coordinates \((\mathrm{lat}, \mathrm{lon})\). To embed these observations into a one-dimensional nonparametric regression framework, we adopt the longitude as the covariate
\[
X := \mathrm{lon},
\]
and treat latitude as the response
\[
Y := \mathrm{lat}.
\]

Although this construction differs from the theoretical parameterization \(X=s\) by arc length, it provides a practical one-dimensional ordering of the data while preserving the relative geometry of the route.

The baseline is the real railway route from Vancouver to Toronto, called ``The canadaian''. The trajectory points of this baseline are obtained from VIA Rail, \url{https://www.viarail.ca/en/developer-resources} which is licensed under Open Government Licence – Canada version 2, \url{https://open.canada.ca/en/open-government-licence-canada}. The existing railway network does not pass through Banff, a region of high scenic significance.
To illustrate the flexibility of the proposed method, we construct a hypothetical railway alignment that incorporates Banff as a fixed waypoint.

\begin{figure}[htbp]
    \centering
    \includegraphics[width=\linewidth, trim=0cm 0cm 0cm 0.7cm, clip]{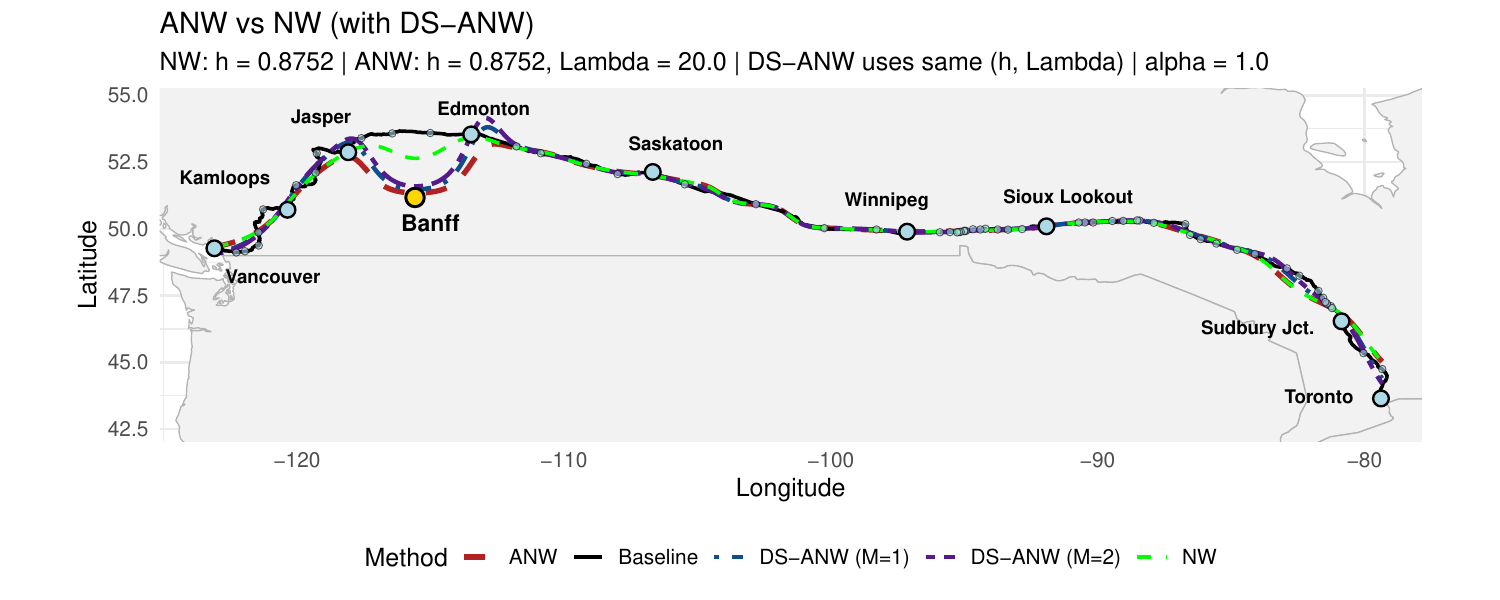}
    \caption{Comparison of NW, ANW, and DS--ANW for one-dimensional railway alignment with a fixed waypoint}

    \label{fig:railway1}
\end{figure}

In Fig.~\ref{fig:railway1}, the fixed waypoint at Banff is enforced exactly by ANW and its data-sharpened variants.
Away from the constraint, ANW produces a trajectory that is largely comparable to NW, while retaining the ability to pull the fit toward the prescribed waypoint.
We also report DS--ANW (with $M=1$ and $M=2$), which further sharpens local features relative to ANW.
However, around Edmonton, where the stop locations become very sparse, DS--ANW may exhibit a noticeable local overshoot: the limited nearby support reduces the stability of kernel-based estimates, and the sharpening update can amplify this local instability.

\begin{figure}[htbp]
    \centering
    \includegraphics[width=\linewidth, trim=0cm 0cm 0cm 1.3cm, clip]{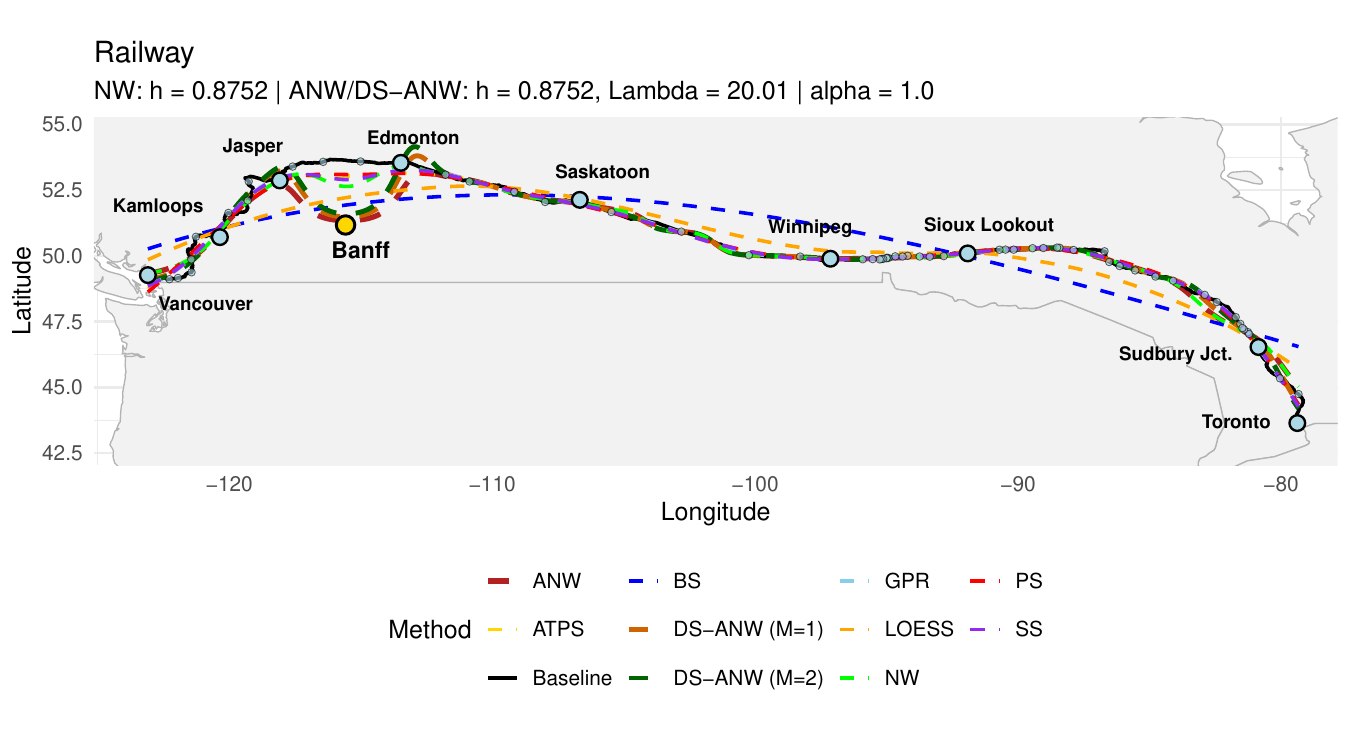}
    \caption{Comparison of nonparametric regression methods for one-dimensional railway alignment under a fixed waypoint constraint}

    \label{fig:railway_comparison}
\end{figure}

In Fig.~\ref{fig:railway_comparison}, we compare ANW and its data-sharpened variants
with six representative nonparametric smoothing methods.
While ANW and DS--ANW are able to pull the fitted trajectory toward the fixed waypoint at Banff, the competing methods remain largely insensitive to this constraint and fail to approach the prescribed location.

\begin{table}[htbp]
\centering
\caption{Performance metrics for the railway application}
\label{tb:railway_metric}
\begin{tabular}{lcccc}
\hline
Model & RMSE & Waypoint Error & Smoothness & CSS \\
\hline
NW               
& 0.29 & 1.47 & $9.57\times10^{-8}$ & 8.00 \\

ANW              
& 0.65 & 0.17 & $2.01\times10^{-7}$ & 3.29 \\

DS--ANW ($M=1$)  
& 0.55 & 0.30 & $3.69\times10^{-7}$ & 3.55 \\

DS--ANW ($M=2$)  
& 0.51 & 0.42 & $4.65\times10^{-7}$ & 3.98 \\

BS (B-spline)    
& 1.07 & 0.91 & $1.86\times10^{-10}$ & 9.10 \\

P-spline         
& 0.22 & 1.92 & $1.02\times10^{-8}$ & 9.92 \\

LOESS            
& 0.70 & 1.04 & $1.65\times10^{-9}$ & 7.87 \\

Smoothing Spline 
& 0.22 & 1.73 & $2.20\times10^{-8}$ & 8.93 \\

GPR              
& 0.22 & 1.73 & $2.39\times10^{-8}$ & 8.98 \\

ATPS             
& 0.22 & 1.86 & $1.34\times10^{-8}$ & 9.62 \\
\hline
\end{tabular}
\end{table}

Table~\ref{tb:railway_metric} reports the performance metrics for the railway alignment application. Compared with unconstrained smoothers, the standard NW estimator attains a relatively low RMSE but exhibits very large waypoint errors, indicating substantial deviations from the prescribed fixed waypoints. Similar behavior is observed for penalized and probabilistic smoothers, including P-splines, smoothing splines, Gaussian process regression (GPR), and adaptive thin-plate splines (ATPS), which prioritize global smoothness at the expense of waypoint adherence.

In contrast, ANW achieves a pronounced reduction in waypoint error, leading to the smallest composite score (CSS) among all competing methods. Although ANW incurs a higher RMSE and reduced smoothness relative to purely global smoothers, its adaptive reweighting mechanism effectively enforces the fixed waypoints, which is critical in the railway setting where fixed waypoints represent mandatory design requirements rather than soft preferences.

The data-sharpened ANW estimators (DS--ANW with $M=1$ and $M=2$) further reduce RMSE compared to ANW, reflecting partial bias correction through iterative sharpening. However, this improvement in global accuracy is accompanied by a noticeable increase in waypoint error and a progressive loss of smoothness as the number of sharpening iterations increases. Consequently, DS--ANW yields larger CSS values than ANW in this application.

This behavior can be attributed to the spatially irregular distribution of railway stations, particularly in sparse regions where local information is limited. In such areas, iterative sharpening may amplify local extrapolation effects, leading to shape distortions and reduced waypoint fidelity. Overall, these results indicate that while DS--ANW can improve global fit, ANW remains the most reliable choice for enforcing strict fixed waypoints in real-world railway alignment problems.

In this application, Banff is imposed as a fixed waypoint to demonstrate the
practical flexibility of the proposed method. In reality, however, the main
transcontinental passenger railway operated by VIA Rail does not serve Banff
directly. Instead, long--distance routes bypass the area in favor of more
northerly corridors with higher population density and operational demand.
Direct rail access to Banff is largely limited to seasonal or luxury tourist
services rather than regular intercity transport. This reflects a real--world
trade--off between scenic value and economic viability, and explains why the
hypothetical alignment constructed here deviates from the existing railway
network.

\subsection{Highway alignment design}

We select the Saskatchewan highway11 which is a major north–south highway in Saskatchewan. The data is from Saskatchewan Government, \url{https://geohub.saskatchewan.ca/datasets/da074f6eb1814ef1b033b7a090c93cd3/about}, under license GOS Standard Unrestricted Use Data Licence v2.0. The dataset consists of 55 major intersections along Highway~11, obtained from Wikipedia, \url{https://en.wikipedia.org/wiki/Saskatchewan_Highway_11}.
To facilitate modeling, the coordinate system is rotated to align the primary direction of the highway with the horizontal axis; the fitted curve is subsequently rotated back to the original orientation.
Hawarden and Martensville are treated as fixed waypoints, as they are well-populated communities located along the Highway~11 corridor.

In the Highway~11 application, the raw geographic coordinates \((x,y) = (\mathrm{lon}, \mathrm{lat})\) form a nearly north--south oriented alignment. In the original coordinate system, this produces an almost vertical curve in the \((x,y)\)-plane, which cannot be represented as a single-valued function \(x = f(y)\) or \(y = g(x)\). Consequently, a direct nonparametric regression is not feasible. To obtain a one-dimensional representation suitable for kernel regression, we apply a rotation of the coordinate system. Let
\[
\begin{pmatrix}
x' \\[3pt]
y'
\end{pmatrix}
=
\begin{pmatrix}
\cos\theta & -\sin\theta \\
\sin\theta & \cos\theta
\end{pmatrix}
\begin{pmatrix}
x \\[3pt]
y
\end{pmatrix},
\qquad \theta = 90^\circ.
\]

This transformation maps the original north--south alignment into an approximately east--west oriented curve. In the rotated system, the transformed alignment becomes nearly horizontal, so that \(y'\) is a approximately a single-valued function of \(x'\):
\[
y' \approx f(x').
\]

This geometric reparameterization greatly simplifies the modeling: the covariate \(X\) can be taken as the horizontal coordinate \(x'\), and the response \(Y\) as the vertical coordinate \(y'\).

\begin{figure}[htbp]
    \centering
    \includegraphics[width=\linewidth, trim=0cm 0cm 0cm 0.7cm, clip]{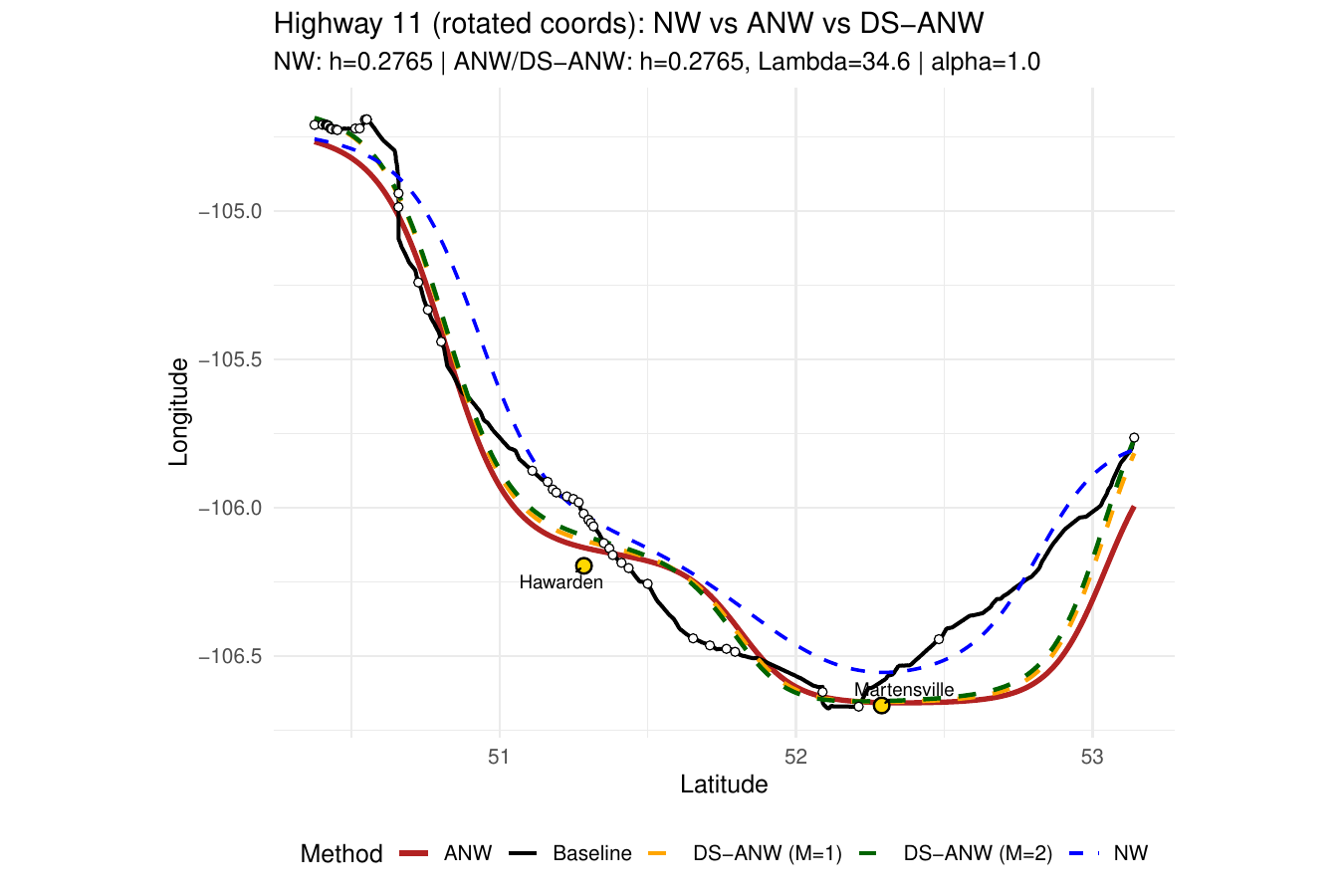}
    \caption{ANW and DS--ANW fitting in the rotated coordinate system for Highway~11 with fixed waypoints}

    \label{fig:highwayfit}
\end{figure}

\begin{figure}[htbp]
    \centering
    \includegraphics[width=\linewidth, trim=0cm 0cm 0cm 0.7cm, clip]{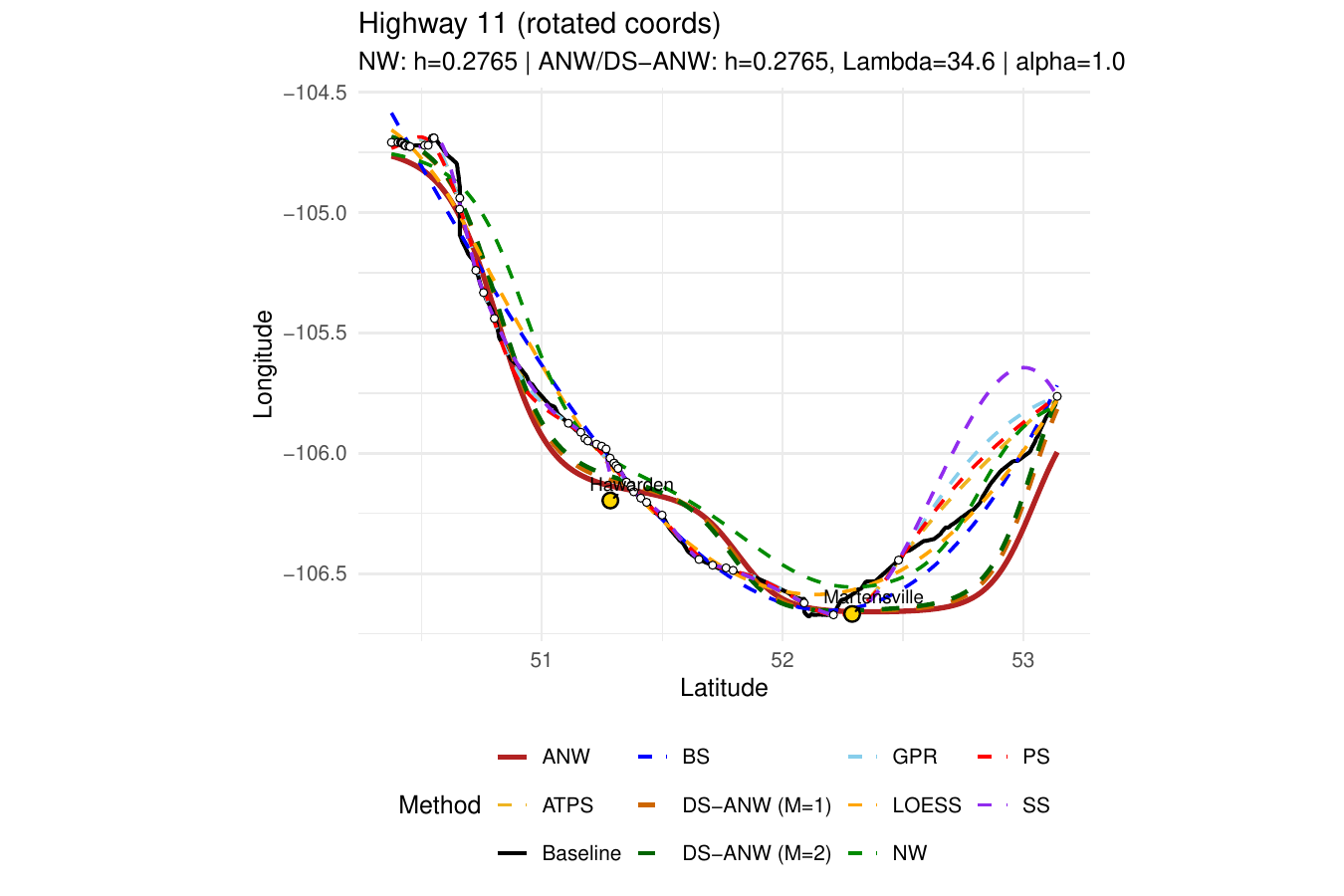}
    \caption{Comparison of nonparametric smoothing methods for Highway~11 in the rotated coordinate system under fixed waypoint constraints}

    \label{fig:highway comparison}
\end{figure}

Fig.~\ref{fig:highwayfit} and Fig.~\ref{fig:highway comparison} present
the fitted alignments in the rotated coordinate system, where the horizontal
axis serves as the covariate and the vertical axis as the response.
This representation allows a direct visual assessment of both waypoint
adherence and global smoothness under different smoothing strategies.

Figure~\ref{fig:highway comparison} compares ANW and DS--ANW with several
representative nonparametric smoothing methods under the same waypoint
constraints. Methods that do not explicitly incorporate waypoint information
tend to exhibit larger deviations near the prescribed locations than ANW-based
approaches, whereas ANW-based methods achieve a clearer balance between
waypoint adherence and global smoothness.

\begin{figure}[htbp]
    \centering
    \includegraphics[width=\linewidth, trim=0cm 1cm 0cm 0.7cm, clip]{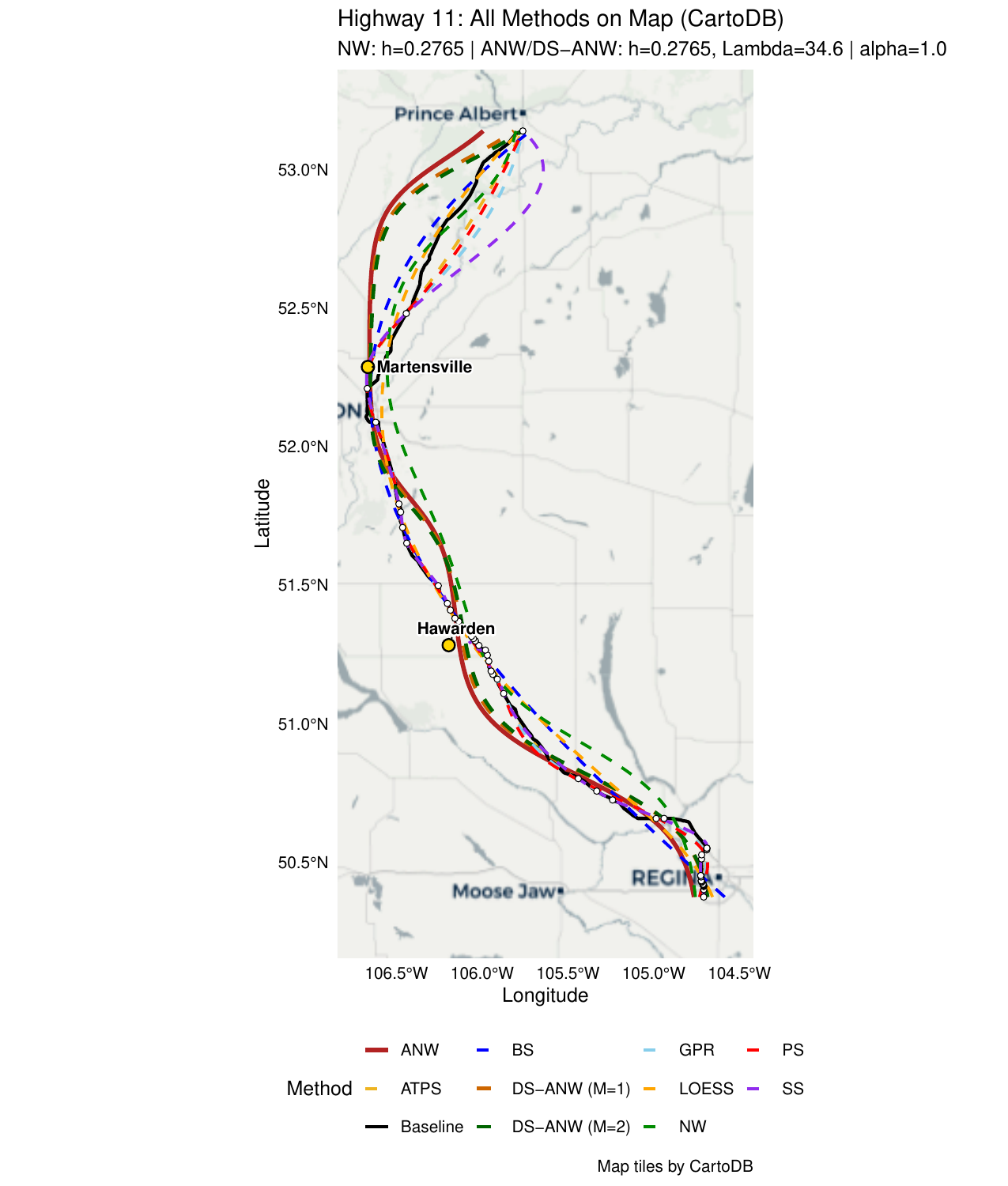}
    \caption{Fitted Highway~11 alignment mapped back to the original geographic coordinates}

    \label{fig:highway}
\end{figure}

Fig.~\ref{fig:highway} maps the fitted alignment back to the original
geographic coordinate system. The fitted curve follows the overall shape of
the highway while remaining close to the fixed waypoints.
The noticeable deviation in a localized latitude range is primarily due to
highly uneven sampling density along the route: in that region, only a small
number of observations are available, providing limited data support for
both NW-type smoothers and waypoint-regularized variants. As a result, the
local fit becomes more sensitive and may drift away from the baseline.

\begin{table}[htbp]
\centering
\caption{Performance metrics for the highway application}
\label{tab:highway_metric}
\begin{tabular}{lcccc}
\hline
Model & RMSE & Waypoint Error & Smoothness & CSS \\
\hline
NW
& 0.14 & 0.15 & $1.92\times10^{-10}$ & 0.58 \\

ANW
& 0.19 & 0.04 & $3.28\times10^{-10}$ & 0.48 \\

DS--ANW ($M=1$)
& 0.16 & 0.06 & $3.85\times10^{-10}$ & 0.52 \\

DS--ANW ($M=2$)
& 0.15 & 0.07 & $3.84\times10^{-10}$ & 0.54 \\

BS (B-spline)
& 0.09 & 1.32 & $2.87\times10^{-11}$ & 0.67 \\

P-spline
& 0.07 & 0.11 & $5.14\times10^{-10}$ & 0.54 \\

LOESS
& 0.07 & 0.12 & $4.65\times10^{-11}$ & 0.48 \\

Smoothing Spline
& 0.15 & 0.08 & $2.56\times10^{-8}$  & 0.64 \\

GPR
& 0.08 & 0.11 & $5.68\times10^{-10}$ & 0.57 \\

ATPS
& 0.06 & 0.11 & $6.03\times10^{-10}$ & 0.54 \\
\hline
\end{tabular}
\end{table}

Table~\ref{tab:highway_metric} summarizes the performance metrics for the highway alignment application. Compared with unconstrained smoothers, the standard NW estimator achieves reasonable global accuracy but exhibits noticeably larger constraint deviations. Several penalized and probabilistic smoothers, including B-splines, P-splines, LOESS, Gaussian process regression (GPR), and adaptive thin-plate splines (ATPS), attain low RMSE values but generally fail to enforce the prescribed fixed waypoints, as reflected by their relatively large waypoint errors.

In contrast, ANW achieves the smallest composite score (CSS) among all methods by substantially reducing the waypoint error while maintaining acceptable global accuracy and smoothness. This indicates that adaptive reweighting of fixed waypoints is particularly effective in the highway setting, where waypoint constraints are moderately strong and spatially well distributed along the route.

The data-sharpened ANW estimators (DS--ANW with $M=1$ and $M=2$) further improve RMSE relative to ANW, reflecting bias reduction through iterative sharpening. However, this improvement in global accuracy is accompanied by a modest increase in waypoint error, and the resulting CSS values remain slightly higher than that of ANW. Notably, the smoothness of DS--ANW remains comparable to ANW, suggesting that mild sharpening does not introduce severe oscillations in this application.

\section{Conclusion}

We proposed an adaptive Nadaraya--Watson estimator (ANW) for route alignment
with fixed waypoints. By introducing waypoint-specific tuning parameters
$\lambda$ while keeping a global bandwidth $h$, ANW decouples constraint
enforcement from smoothing and avoids the local irregularities often produced
by naive bandwidth shrinking near fixed waypoints. We further combined ANW with
iterated data sharpening (IDS2) to obtain DS--ANW, which reduces smoothing bias
without changing the waypoint weighting mechanism.

Our theory establishes pointwise convergence at fixed waypoints under
appropriate $\lambda$ scaling and shows that ANW is asymptotically equivalent
to the classical NW estimator at interior points. Numerical studies in 1D and
2D demonstrate that ANW provides a stable balance between global accuracy,
smoothness, and waypoint adherence, especially when fixed waypoints conflict
with the baseline geometry. DS--ANW can further improve RMSE with a small
number of sharpening steps (typically $M=1$--$2$), but may yield diminishing
returns and can be less stable in sparsely sampled regions. Applications to
railway and highway alignments illustrate that augmenting external design
requirements as deterministic waypoints offers a simple and transparent way to
integrate engineering constraints into a statistical smoothing pipeline.

\section{Discussion: Robustness considerations and possible extensions}

The proposed adaptive Nadaraya--Watson estimator (ANW), together with its
data--sharpened variant DS--ANW, is designed to balance smoothness and strict
waypoint adherence under the classical squared--error regression framework.
As with most kernel--based smoothers, the estimator is optimal under
light--tailed noise assumptions but may become sensitive to outliers or
local contamination, particularly in regions with sparse sampling.

From a practical alignment--design perspective, such contamination may arise
from measurement errors, digitization artifacts, or irregular data acquisition
along the route. In these situations, it is natural to consider robustness
enhancements that limit the influence of extreme observations while preserving
the waypoint attraction mechanism introduced by adaptive weighting.

One possible direction is to incorporate ideas from distributionally robust
optimization (DRO) into the ANW framework \cite{blanchet2019robust}. At a conceptual level, this can be
achieved by replacing the pure squared--error objective with a penalized or
robustified local criterion that shrinks extreme local averages toward a
baseline value. Importantly, such modifications can be implemented without
altering the fixed waypoint tuning parameters $\lambda$, so that waypoint enforcement
remains intact.

However, robustness typically comes at the cost of increased bias, especially
near regions of strong local curvature or rapidly changing geometry. This trade--off
is particularly relevant in alignment design problems, where both geometric
fidelity and waypoint accuracy are critical. While robustification may suppress
the effect of outliers, it can also dampen legitimate local features of the
trajectory.

Data sharpening provides a complementary mechanism in this context. As a
bias--reduction wrapper that operates independently of the underlying estimator,
it can, in principle, be applied on top of both standard and robustified ANW
variants. This suggests a flexible modular framework in which waypoint
enforcement, robustness, and bias reduction can be combined in different ways,
depending on the characteristics of the data and the design requirements.

A systematic investigation of distributionally robust ANW estimators and their
interaction with data sharpening is beyond the scope of the present paper and
is left for future work. Such an extension would require careful theoretical
analysis of the resulting bias--variance trade--offs, as well as extensive
empirical evaluation under heavy--tailed or contaminated noise models.

\section{Data Availability}
The dataset can be requested by emailing xiaoping.shi@ubc.ca.

\section{Acknowledgments}

The author Qiong Li was supported by NSFC (Grant
No. 12271047) and Guangdong and Hong Kong Universities “1+1+1” Joint Research Collaboration Scheme (2025A0505000010). The author Xiaoping Shi was supported by the NSERC Discovery Grant RGPIN 2022-03264,  the NSERC Alliance International Catalyst Grant ALLRP 590341-23, and the University of British Columbia Okanagan (UBC-O) Vice Principal Research in collaboration with the UBC-O Irving K. Barber Faculty of Science.

\section{Author contributions}
S.D. and X.S. designed this research. S.D. performed this research and wrote the main manuscript. Y.C. assisted in performing the comparative analysis. W.Y., Q.L., and X.S. reviewed and revised the manuscript. All authors reviewed the manuscript. 

%
%
\bibliography{main}

\end{document}